\documentclass[letterpaper,english,american]{article}
\usepackage[utf8]{inputenc}
\usepackage{color}
\usepackage{babel}
\usepackage{prettyref}
\usepackage{mathtools}
\usepackage{amsmath}
\usepackage{amsthm}
\usepackage{amssymb}
\usepackage{graphicx}
\usepackage{microtype}
\usepackage[bookmarks=false,
 breaklinks=false,pdfborder={0 0 1},backref=section,colorlinks=false]
 {hyperref}

\makeatletter

\pdfpageheight\paperheight
\pdfpagewidth\paperwidth

\usepackage{empheq}

\usepackage[accepted]{./icml2026}

\usepackage{amsthm}
\usepackage{cuted}

\usepackage[capitalize,noabbrev]{cleveref}
\usepackage{physics}
\usepackage{cuted}

\theoremstyle{plain}

\theoremstyle{definition}
\theoremstyle{remark}

\newrefformat{cap}{\hyperref[#1]{Fig.~\ref{#1}}}
\newrefformat{fig}{\hyperref[#1]{Fig.~\ref{#1}}}
\newrefformat{tab}{\hyperref[#1]{Table ~\ref{#1}}}
\newrefformat{sec}{\hyperref[#1]{Section~\ref{#1}}}
\newrefformat{subsec}{\hyperref[#1]{Section~\ref{#1}}}
\newrefformat{sub}{\hyperref[#1]{Section~\ref{#1}}}
\newrefformat{cha}{\hyperref[#1]{Chapter~\ref{#1}}}
\newrefformat{app}{\hyperref[#1]{Appendix~\ref{#1}}}

\ifdefined\showcaptionsetup
 \PassOptionsToPackage{caption=false}{subfig}
\fi
\usepackage{subfig}
\makeatother

\begin{document}
\global\long\def\bR{\mathbb{R}}%
\global\long\def\llangle{\langle\!\langle}%
\global\long\def\rrangle{\rangle\!\rangle}%
\global\long\def\T{\mathrm{\mathrm{\top}}}%
\global\long\def\erf{\mathrm{erf}}%
\global\long\def\tf{\tilde{f}}%
\global\long\def\N{\mathcal{N}}%
\global\long\def\cD{\mathcal{D}}%
\global\long\def\bI{\mathbb{I}}%
\global\long\def\tr{\mathrm{tr}}%
\global\long\def\tQ{\tilde{Q}}%
\global\long\def\tQ{\tilde{Q}}%
\global\long\def\order{\mathcal{O}}%
\global\long\def\const{\mathrm{const.}}%
\global\long\def\cW{\mathcal{W}}%
\global\long\def\cQ{\mathcal{Q}}%
\global\long\def\tU{\tilde{U}}%
\global\long\def\dd{\mathrm{d}}%
\global\long\def\tv{\tilde{v}}%
\global\long\def\tC{\tilde{C}}%
\icmltitlerunning{Dynamics of neural scaling laws in random feature regression with powerlaw-distributed kernel eigenvalues} 

\twocolumn[
\icmltitle{Dynamics of neural scaling laws in random feature regression with powerlaw-distributed kernel eigenvalues}

\icmlsetsymbol{equal}{*}

\begin{icmlauthorlist}

\icmlauthor{Jakob Kramp}{equal,fzj,rwth_phd} 
\icmlauthor{Javed Lindner}{equal,fzj,rwth_phd}
\icmlauthor{Moritz Helias}{equal,fzj,rwth}
\end{icmlauthorlist}

\icmlaffiliation{fzj}{
Institute for Advanced Simulation (IAS-6), Computational and Systems Neuroscience, J\"ulich Research Centre, J\"ulich, Germany}
\icmlaffiliation{rwth}{Department of Physics, RWTH Aachen University, Aachen, Germany}
\icmlaffiliation{rwth_phd}{RWTH Aachen University, Aachen, Germany}

\icmlcorrespondingauthor{Jakob Kramp}{jakob.kramp@rwth-aachen.de}

\icmlkeywords{Keyword}

\vskip 0.3in
]
\printAffiliationsAndNotice{\icmlEqualContribution}
\begin{abstract}
\foreignlanguage{english}{Training large neural networks exposes
neural scaling laws for the generalization error, which points to
a universal behavior across network architectures of learning in high
dimensions. It was also shown that this effect persists in the limit
of highly overparametrized networks as well as the Neural network
Gaussian process limit. We here develop a principled understanding
of the typical behavior of generalization in Neural Network Gaussian
process regression dynamics. We derive a dynamical mean-field theory
that captures the typical case learning dynamics: This allows us to
unify multiple existing regimes of learning studied in the current
literature, namely Bayesian inference on Gaussian processes, gradient
flow with or without weight-decay, and stochastic Langevin training
dynamics. Employing tools from statistical physics, the unified framework
we derive in either of these cases yields an effective description
of the high-dimensional microscopic behavior of networks dynamics
in terms of lower dimensional order parameters. We show that collective
training dynamics may be separated into the dynamics of $N$ independent
eigenmodes, whose evolution equations are only coupled through collective
response functions and a common statistics of an effective, independent
noise.  Our approach allows us to quantitatively explain the dynamics
of the generalization error by linking spectral and dynamical properties
of learning on data with power law spectra, including phenomena such
as neural scaling laws and the effect of early stopping. }
\end{abstract}

\section{Introduction }

\selectlanguage{english}%

Natural data, such as language and image data, ubiquitously shows
power law distributions in their principal component spectra \cite{vanderschaafModellingPowerSpectra1996,koch1F2Characteristics2010,bulanadiIdentifyingAnalyzingPowerlaw2024}.
This property persists even if the data is mapped into a feature space.
Training large neural networks on such data exposes the phenomenon
of neural scaling laws for the generalization error, which points
to a universal behavior of learning in high dimensions. Such behavior
is observed across neural architectures \cite{Kaplan20_arxiv}. The
relevance of this universal behavior also has practical applications,
as it provides a theory based approach to use smaller and computationally
more efficient models to gauge the behavior of bigger models without
running costly training. It was also shown that this universality
persists in the limit of large numbers of neurons, which reduces networks
to Gaussian process regression \cite{neural_jacot_2018,Lee18}. Gaussian
process regression, and thus also linear regression, offers an analytically
tractable setting that captures key qualitative behaviors of real-world
overparameterized networks, including scaling laws and the fabled
double descent phenomenon \cite{Nakkiran_2021}. We thus focus our
work on a principled understanding of the typical behavior of generalization
in linear regression dynamics.

To this end, we derive a dynamical mean-field theory that captures
the typical learning dynamics in the limit of large numbers of neurons
and training samples. This approach allows us to unify multiple existing
approaches to the study of learning found in the current literature,
namely Bayesian inference on Gaussian processes, gradient flow with
or without weight-decay, and stochastic Langevin training dynamics.
We arrive at this versatile theory by employing statistical field
theory of disordered systems. As in statistical physics, the high-dimensional
microscopic behavior of the network's dynamics can be effectively
described by a set of interpretable lower dimensional order parameters.
The theory shows that the collective training dynamics can be separated
into the dynamics of $N$ independent eigenmodes, whose evolution
equations are only coupled through a collective response function
and the jointly determined statistics of an effective noise. This
approach enables us to link the success of machine learning heuristics
like ``early-stopping'' (see e.g. \foreignlanguage{american}{\cite{advaniHighdimensionalDynamicsGeneralization2020}})
 to the spectral and dynamical properties of learning on power law
distributed data. Specifically, we
\begin{itemize}
\item derive an effective mean-field theory that quantitatively explains
the dynamics of the generalization error, including phenomena such
as neural scaling laws and the effect of early stopping,
\item explain how the analytically found, disorder induced, effective self-coupling
slows down the learning dynamics,
\item and obtain analytical results that relate the power-law exponent of
the feature kernel, regularization, and early stopping time to obtain
a minimal generalization gap.
\end{itemize}

\selectlanguage{american}%

\section{Related works}

\selectlanguage{english}%
Past work has focused on different aspects of learning that we unify
here: \citet{Krogh_Hertz_1990,generalization_krogh_1992} and \citet{learning_krogh_1992}
were the first to consider the generalization properties of linear
regression, its learning dynamics in a stochastic setup and the influence
of data and network properties on learning dynamics as well as dynamics
close to the interpolation threshold. The setup is close to ours,
but differs by them not considering  kernel regression with power-law
distributed eigenvalues. \citet*{Dunmur_1993} studied the effects
of noisy data on the performance of linear systems in a stochastic
setup at infinite training time $t\rightarrow\infty$. Further approaches
include computing the asymptotic limit at $t\rightarrow\infty$ utilizing
Green's function approaches \cite{Sollich_1994}, where \citet{Halkjaer_Winther_1996}
also considered the effects of modes of the input covariance on the
convergence of gradient descent. 

More recent literature is structured into static and dynamic approaches
in the following. \\
Among the static approaches, which is the limit of infinite training
time, our work is methodologically closest related to \citet{spectral_canatar_2021}
and \citet{spectrum_bordelon_2020}, who consider feature regression
in a reproducing kernel Hilbert space with power-law distributed kernel
eigenvalues. While technically the authors introduce stochasticity
(finite temperature), they do not exploit the link to Bayesian inference
which we expose here. \citet*{cui_2022} extend \citet{spectrum_bordelon_2020}
by adding label noise in their setup. \citet*{spigler2020asymptotic}
compare student-teacher kernel regression to soft SVM classification
and derive generalization error exponents from smoothness and dimensionality
assumptions. \citet*{solvable_maloney_2022} investigated neural scaling
laws in linear regression with power-law distributed eigenvalue spectra
and L2-regularization. \citet*{defilippis2024dimension} study static
random feature regression without specific scaling assumptions on
sample or system size. \citet*{scaling_atanasov_2024} derive deterministic
equivalents for static random feature regression via the S-transform,
compactly charting scaling law exponents across many settings. 

On the side of the dynamical approaches, \citet{advaniHighdimensionalDynamicsGeneralization2020}
studied the dynamics of deterministic gradient flow in linear regression
with L2 regularization; compared to our work, their work lacks temporal
stochasticity and power-law distributed eigenvalue spectra; instead
their data is Gaussian i.i.d. \citet*{bordelonDynamicalModelNeural2024a}
studied learning dynamics as gradient flow with power law data in
a type of linear random feature models, which showed how effects
of feature learning may affect neural scaling laws, while considering
systems in a gradient flow regime. Our setup is most closely related
to theirs, however, they do not cover the effects of L2 regularization
and in addition our work requires fewer and more interpretable order
parameters. While having access to the dynamics, the authors do not
investigate the response functions and the resulting slowdown of the
training dynamics further, but rather focus on the time dynamics of
the loss. \citet*{paquette2024phases} study the compute-optimal
Pareto frontier in random feature regression trained by SGD with finite
batch size and power-law features, deriving a deterministic Volterra
equation via random matrix theory and identify multiple scaling phases.
Like \citet{bordelonDynamicalModelNeural2024a}, they use a down-projected
student-teacher setup. We instead study Langevin gradient flow with
identical student and teacher features, which, beyond feature regression,
naturally connects to the Bayesian posterior. 
\selectlanguage{american}%

\section{Setup}

\foreignlanguage{english}{ We consider a teacher-student setup \cite{generalization_krogh_1992}
with}
\begin{align}
f_{\mu}= & w^{\T}\psi_{\mu}\,,\\
y_{\mu}= & \overline{w}^{\top}\psi_{\mu}\,,\label{eq:label_def}
\end{align}
\foreignlanguage{english}{where $\psi(x_{\mu})\eqqcolon\psi_{\mu}\in\mathbb{R}^{N}$
are considered the features of the data $x_{\mu}$ and $\mu$ indexes
the training points. The labels $y_{\mu}$ are generated by the teacher
weights $\overline{w}\in\mathbb{R}^{N}$ while $w\in\mathbb{R}^{N}$
are the student weights our model aims to learn. While we consider
a noiseless teacher here, label noise can easily be included in the
theory as demonstrated in \prettyref{sec:label-noise}. The train
set for this regression task is given by $\mathcal{D}=\{(x_{\mu},y_{\mu})\}_{\mu=1,\ldots,P}$
and $(x_{*},y_{*})\notin\mathcal{D}$ is a test point. We assume $x_{\mu},x_{\ast}\stackrel{\text{i.i.d.}}{\sim}p(x)$
to be drawn from the same distribution.  We solve the regression
task on the teacher-student setup with an L2-regularized squared loss
\begin{equation}
H(w,\mathcal{D}):=\frac{1}{2}\,\sum_{\mu=1}^{P}\big(y_{\mu}-f_{\mu}(w)\big)^{2}+\frac{1}{2g\beta}\sum_{i=1}^{N}w_{i}^{2}\quad.\label{eq:def_H}
\end{equation}
Instead of considering gradient flow, which would yield a single maximum
a posteriori point estimate for the weights and the network output,
we consider a setting where the loss is minimized under stochastic
Langevin gradient descent, which has been proven to serve as a good
approximation to more realistic training scenarios (see e.g. \cite{naveh20_01190}).
The update equation for the weights reads}
\begin{align}
\frac{\partial}{\partial t}w_{i}(t) & =-\partial_{w_{i}}H\big(w(t),\mathcal{D}\big)\,+\zeta_{i}(t)\,,\\
\langle\zeta_{i}(t)\zeta_{j}(s)\rangle_{\zeta} & =2/\beta\,\delta_{ij}\,\delta(t-s)\quad.
\end{align}
\foreignlanguage{english}{This setup for regression is general in
the sense that it covers three aspects: Firstly, $\beta\to\infty$
yields gradient flow with weight decay ($1/(2g\beta)\neq0$ finite)
or without weight decay ($1/(2g\beta)=0$).}

\selectlanguage{english}%
Secondly, the limit $t\to\infty$ and $\beta$ finite yields the stationary
distribution of the weights $p(w\vert\mathcal{D})\propto\exp\big(-\beta\,H(w,\mathcal{D})\big)$
\cite{statistical_seung_1992,naveh20_01190}, equivalent to the Bayesian
posterior for the Gaussian process $f(x)\sim\mathcal{N}(0|\kappa+\beta^{-1}\mathbb{I})$
with kernel 
\begin{equation}
\kappa(x,x^{\prime})=g\,\psi(x)^{\top}\psi(x^{\prime})\label{eq:kernel_kappa}
\end{equation}
 and noisy training samples $y_{\mu}+\epsilon_{\mu}$ with Gaussian
noise $\epsilon_{\mu}\stackrel{\text{i.i.d.}}{\sim}\mathcal{N}(0,\beta^{-1})$
(see \prettyref{sec:Bayesian-interpretation-of} for details). And
thirdly, considering the dynamics of linear regression is a starting
point to study real networks in their lazy learning limits, such as
the neural network Gaussian process \cite{Neal1995BayesianLF,computing_williams_1996},
the neural tangent kernel \cite{neural_jacot_2018}, and random feature
models \cite{Rahimi2007}, which are all linearized forms of real
networks.

In addition to the dynamics we are particularly interested in generalization
error $\mathcal{L}_{\mathrm{test}}$. It is expected to show little
variation across different drawings of train patterns. We will therefore
be interested in the average $\langle\mathcal{L}_{\mathrm{test}}\rangle_{\cD}$
over many such drawings to describe the generalization properties
of the system. The test error measures the error of the learned predictor
$f_{*}$ made on a new drawing of the data $x_{\ast}\sim p,y_{\ast}=\bar{w}^{\top}x_{\ast}$,
which hence is
\begin{align}
\langle\mathcal{L}_{\mathrm{test}}\rangle & =\frac{1}{2}\,\big\langle(y_{\ast}-f_{\ast})^{2}\big\rangle_{x_{\ast}\sim p(x)\,,y_{\ast}=\bar{w}\,x_{\ast}}\,.\label{eq:E_test}
\end{align}

\selectlanguage{american}%

\section{Theory}

\subsection{Background}

\selectlanguage{english}%
Following along the lines of \cite{spectral_canatar_2021} we assume
that the features of the inputs fulfill the orthogonality relation\foreignlanguage{american}{
\begin{equation}
\delta_{ij}\,\eta_{i}=\int\psi_{i}(x)\psi_{j}(x)\,p(x)\,dx\,.
\end{equation}
}A common example on a finite set of data points are the principal
components of the empirical covariance matrix. Further we assume a
Gaussian distribution on the features $\psi_{i}\sim\mathcal{N}(0,\eta_{i}\delta_{ij})$
which are i.i.d. across different training samples $\mu$ for different
draws $\psi(x_{\mu})$ themselves. While the theory allows for arbitrary
$\eta_{i},$ we assume $\eta_{i}$ to show a power law decay 
\begin{align}
\eta_{i} & =\eta_{1}\,i^{-1-\gamma}\,\label{eq:power_law}
\end{align}
from here on. Together with the definition of the teacher-student
model we can rewrite our training objective \eqref{eq:def_H} as
\begin{align}
H & :=\frac{1}{2}\sum_{\mu=1}^{P}\left[\sum_{i=1}^{N}(\overline{w}_{i}-w_{i})\psi_{i}(x_{\mu})\right]^{2}+\frac{1}{2g\beta}\sum_{i=1}^{N}w_{i}^{2}\label{eq:H_in_modes}\\
 & =\frac{1}{2}\sum_{i,j=1}^{N}(\overline{w}_{i}-w_{i})Q_{ij}(\overline{w}_{j}-w_{j})\\
 & +\frac{1}{2g\beta}\sum_{i=1}^{N}w_{i}^{2}+\const\,,
\end{align}
where we dropped terms that are independent of $w$ and defined the
feature kernel $Q_{ij}\in\mathbb{R}^{N\times N}$ as
\begin{equation}
Q_{ij}=\sum_{\mu=1}^{P}\psi_{i}(x_{\mu})\psi_{j}(x_{\mu})\quad.\label{eq:Q_definition}
\end{equation}
Another view on this setting is to consider $\psi$ the eigenmodes
of the kernel that describes the Gaussian limit of a neuronal network's
output on which the final layer of weights $w$ is trained. The loss
\eqref{eq:H_in_modes} implies the gradient dynamics for the coefficients
$w_{i}$\foreignlanguage{american}{
\begin{align}
\frac{\partial}{\partial t}w_{i}(t) & =\sum_{j=1}^{N}\,Q_{ij}(\overline{w}_{j}-w_{j})-\frac{1}{g\beta}w_{i}+\zeta_{i}(t)\,,\label{eq:eom_w_main}
\end{align}
The weights are hence coupled by the matrix $Q$. We will find below
that in the limit of high dimensions, we achieve a partial decoupling
of the statistics of the modes.}

Considering the structure of the right hand side we define $v_{i}:=\overline{w}_{i}-w_{i}$
which is the discrepancy between the teacher and the student weight
as well as $\Omega_{ij}:=Q_{ij}+\delta_{ij}/(g\beta)$ and rewrite
the gradient descent as
\begin{equation}
\frac{\partial}{\partial t}v_{i}(t)=-\sum_{j=1}^{N}\Omega_{ij}v_{j}(t)+\frac{1}{g\beta}\overline{w}_{i}+\zeta_{i}(t)\quad.\label{eq:of_motion_single}
\end{equation}

\selectlanguage{american}%
To investigate the distribution on $v_{i}(t)$, we utilize the MSRDJ
formalism \cite{Martin73,dedominicis1976_247,janssen1976_377} following
the lines of \cite{helias20_970} to state a moment-generating functional
$Z$ of the stochastic differential equation
\begin{align}
Z & =\int\mathcal{D}v\,\mathcal{D}\tilde{v}\,\exp\big(S(v,\tilde{v})\big)\\
S(v,\tilde{v}) & =\int dt\sum_{i=1}^{N}\tilde{v}_{i}(t)\,\Big(\partial_{t}v_{i}(t)+\sum_{j=1}^{N}\Omega_{ij}v_{j}(t)-\frac{\overline{w}_{i}}{g\beta}\,\Big)\\
 & +\beta^{-1}\sum_{i=1}^{N}\tilde{v}_{i}(t)^{2}\,.\nonumber 
\end{align}
Ultimately, we are interested in the typical behavior of the system
which is invariant under different redraws of the training data set.
Our analysis shows that the training process is indeed self-averaging,
which means that the dynamics of macroscopic observables such as the
test loss, in the limit of sufficiently large systems and data samples,
becomes sharply peaked on its typical behavior. Technically, this
allows one to consider the average of the moment-generating functional
over different draws of the data, which enters in terms of the features
in the matrix $Q$ and hence implies the averaged moment generating
function
\[
\langle Z\rangle_{Q}=\int\mathcal{D}v\,\mathcal{D}\tilde{v}\,\big\langle\exp\big(S(v,\tilde{v})\big)\big\rangle_{Q}\,.
\]
We here use that the normalization of the dynamical generating functional
$Z$ is independent of $Q$, which allows us to directly average $Z$
(as opposed to $\ln Z$ in the static case) \cite{dedominicisDynamicsSubstituteReplicas1978,helias20_970}.
By the assumption of Gaussian features $\psi$ in \eqref{eq:Q_definition},
the feature matrix $Q$ is a Wishart matrix. After taking the disorder
average, the system can be described with a new action $\bar{S}$
as 
\[
\expval Z=\int\mathcal{D}v\mathcal{D}\tilde{v}\,\exp\big(\bar{S}(v,\tilde{v})\big)\quad,
\]
whose detailed form is derived in \prettyref{sec:Generating-functional-in}.
As the MSRDJ formalism gives us a mapping of an equation of motion
to a Gaussian action, the aim is to recover an effective equation
of motion from the disorder averaged action $\bar{S}.$ To this end,
we approximate the distribution $e^{\bar{S}}$ by the Gaussian distribution
$\N\big[(\bar{v},\bar{\tilde{v}}),G\big]$ that maximizes the negative
Kullback-Leibler (KL) divergence
\begin{align}
\Gamma(\bar{v},\bar{\tilde{v}},G) & :=-\mathrm{KL}(\N\big[(\bar{v},\bar{\tilde{v}}),G\big]\|e^{\bar{S}(v,\tv)})\label{eq:KL_divergence}\\
 & =\big\langle S(v,\tv)\big\rangle_{(v,\tv)\sim\N\big[(\bar{v},\bar{\tilde{v}}),G\big]}+\frac{1}{2}\ln\det\big[G\big]\quad.\nonumber 
\end{align}
The details of this variational Gaussian approximation (VGA) are given
in \prettyref{app:Identification-of-order-papameters} and \prettyref{app:VGA}.
We find that the stationary points of the KL divergence with regard
to the parameters $\bar{v},$ $\bar{\tv}$, and $G$ can be expressed
through a set of order parameters
\begin{align}
R(t,s) & =\sum_{k}\eta_{k}v_{k}(t)\tilde{v}_{k}(s)\,,\nonumber \\
\tilde{C}(t,s) & =\sum_{k}\eta_{k}\tilde{v}_{k}(t)\tilde{v}_{k}(s)\,,\label{eq:def_order_params}\\
C(t,s) & =\sum_{k}\eta_{k}v_{k}(t)v_{k}(s)\,,\nonumber 
\end{align}
that concentrate around their expectation values as $N\rightarrow\infty$.

\subsection{Equations of motion for the first and second order statistics\protect\label{sec:Equations-of-motion}}

\begin{figure}
\includegraphics[width=0.95\columnwidth]{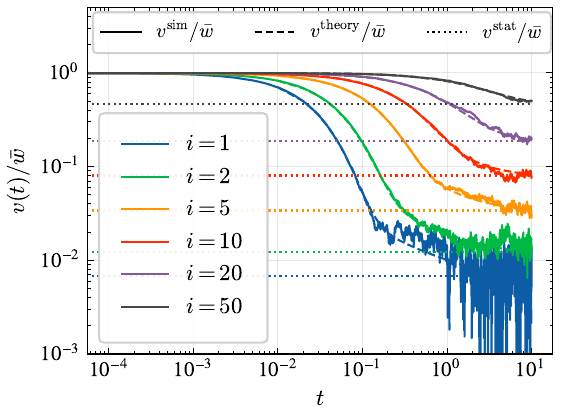}\caption{Time evolution of normalized mean discrepancy $v_{i}(t)/\bar{w}_{i}=(\bar{w}_{i}-w_{i}(t))/\bar{w}_{i}$.
Simulation (full curves) compared to theory (dashed curves, \eqref{eq:effective_eom-1}).
Different curves show different modes $\eta_{i}$ from blue (large
$\eta_{i}$) to black (small $\eta_{i}$, see legend).\protect\label{fig:Time-evolution-of-discrepancy}\protect\label{fig:v_bar}}
\end{figure}

Computing the stationary point of the KL divergence, we obtain a
set of $N$ effective equations of motion that are decoupled across
features
\begin{align}
\left[\partial_{t}+\frac{1}{g\beta}\right]v_{i}(t)+P\eta_{i}\int K(t-s)v_{i}(s)\dd s & =\frac{\bar{w_{i}}}{g\beta}+\xi_{i}(t),\label{eq:effective_eom}
\end{align}
with $i=1,\ldots,N$ and where we assume the initial condition $w_{i}(0)=0$,
which corresponds to $v_{i}(0)=\bar{w}_{i}$. Despite the processes
being statistically independent across features $i$, they interact
indirectly because they are driven by uncorrelated centered Gaussian
noises $\xi_{i}$, which, however, share a common and self-consistently
determined covariance function
\begin{align}
\langle\xi_{i}(t)\xi_{j}(s)\rangle & =\label{eq:xi_xi_correlator-1}\\
 & \delta_{ij}\left[2\beta^{-1}\delta(t-s)+P\eta_{i}\big\{ K\ast\bar{C}\ast K^{\T}\big\}(t,s)\right].
\end{align}
The noise is no longer white, but correlated in time with the correlations
mediated through the mean correlation $\bar{C}:=\expval C$ controlled
by the order parameter \prettyref{eq:def_order_params}. The temporal
memory kernel \textbf{$K(t,s)$}, in addition, mediates a self-coupling
of the process with its own past and obeys a linear convolution equation
\begin{align}
K(t-s) & =\int dt^{\prime}\,\bar{R}(t-t^{\prime})\,K(t^{\prime}-s)+\delta(t-s)\label{eq:K_of_R}
\end{align}
with $\bar{R}:=\expval R$ and $R$ defined in \eqref{eq:def_order_params},
which can be written as the weighted sum
\begin{equation}
\bar{R}(t-s)=\sum_{i}\eta_{i}\,G_{i}(t-s)\label{eq:R_of_G}
\end{equation}
where $G_{i}(t-s):=\expval{v(t)\tv(s)}$ is the Green's function of
\eqref{eq:effective_eom} and obeys itself 
\begin{align}
 & -\left[\partial_{t}+\frac{1}{g\beta}\right]G_{i}(t-s)\label{eq:response_function}\\
 & =P\eta_{i}\!\int\dd u\,K(t-u)\,G_{i}(u-s)+\delta(t-s)\,.\nonumber 
\end{align}
By their linearity and by the appearing convolutions, this set of
equations is conveniently solved in the Fourier domain $t-s\to\omega$
as $\hat{K}(\omega)=\big[1-\hat{R}(\omega)\big]^{-1}\,$, $\hat{R}(\omega)=\sum_{i}\eta_{i}\,\hat{G}_{i}(\omega)\,,$
and 
\begin{equation}
\hat{G}_{i}(\omega)=\text{\ensuremath{-\big[i\omega+\frac{1}{g\beta}+P\eta_{i}\,\hat{K}(\omega)\big]^{-1}}}\,.\label{eq:green_fourier}
\end{equation}

The equation of motion for the mean is obtained by taking the expectation
value over the noise of \eqref{eq:effective_eom}, which yields
\begin{align}
\left[\partial_{t}+\frac{1}{g\beta}\right]\,\langle v_{i}(t)\rangle+P\eta_{i}\,\int\,K(t-s)\,\langle v_{i}(s)\rangle\,ds & =\frac{\bar{w_{i}}}{g\beta}\,,\label{eq:effective_eom-1}
\end{align}
where the initial condition is $\langle v_{i}(0)\rangle=\bar{w}_{i}$.
This expression shows that the time-scale of the $i$-th mode is controlled
by the self-interaction term $P\,\eta_{i}K\ast\langle v\rangle$,
which hence depends inversely on the $i$-th eigenvalue $\eta_{i}$.
This timescale controls how fast the corresponding mode is learned.
Large eigenmodes therefore approach the teacher signal more quickly
than slow ones as can be seen in \prettyref{fig:Time-evolution-of-discrepancy}
where the time evolution of the mean discrepancies for different modes
is displayed in relation to the theoretical mean value $\expval v$
given by \prettyref{eq:effective_eom-1}. Likewise, the amount of
data enters here multiplicatively as $P,$ showing that more data
leads to faster learning of the teacher, as intuitively expected.
The same is true for the stationary value of $\expval v$, which is
also determined through the self-interaction term as
\begin{equation}
\lim_{t\to\infty}\expval v_{i}(t)=\frac{\bar{w}}{1+g\beta P\eta_{i}\int_{0}^{\infty}K(t)\,\dd t}\quad.\label{eq:stationary_value}
\end{equation}
Its final value is thus influenced through the kernel eigenvalue $\eta_{i}$,
that data size $P$, as well as regularization strength $g\beta$
in the same manner. Given the fact that $K$ acts as a mean field
coupling, it however depends by \prettyref{eq:K_of_R} and \prettyref{eq:R_of_G}
on the entire distribution of eigenvalues instead of a single one.

In addition, it can be observed that the fluctuations  of $v$ around
the mean $\langle v(t)\rangle$ increase over time. Their covariance
follows from \eqref{eq:effective_eom} as the Green's function \eqref{eq:response_function}
convolved with the noise $\xi$, which hence yields
\begin{align}
 & \langle\delta v_{i}(t)\delta v_{i}(s)\rangle\nonumber \\
 & =2/\beta\,\int_{0}^{\min(t,s)}du\,G_{i}(t-u)G_{i}(s-u)\\
 & +P\eta_{i}\left[G_{i}*K*\bar{C}*K^{\top}*G_{i}^{\top}\right](t,s)\,,\label{eq:variance_part}
\end{align}
where 
\begin{align*}
 & \left[K*\bar{C}*K^{\top}\right](u,u^{\prime})\\
 & =\int_{0}^{u}dt\int_{0}^{u^{\prime}}ds\,K(u-t)\,\bar{C}(t,s)\,K(u^{\prime}-s)\,,
\end{align*}
which are solved self-consistently with $\bar{C}$, which, in turn,
requires the covariance $\langle\delta v_{i}(t)\delta v_{i}(s)\rangle$
and the mean given by \eqref{eq:eom_mean}. The fluctuations are both
dependent on the initial Langevin noise $\zeta$ and on the correlation,
which leads to a self-reinforcement over time whose consequences are
discussed in the following chapter.

\subsection{Time dynamics of the generalization error}

\begin{figure*}[!t]
\subfloat{\includegraphics[width=0.95\columnwidth]{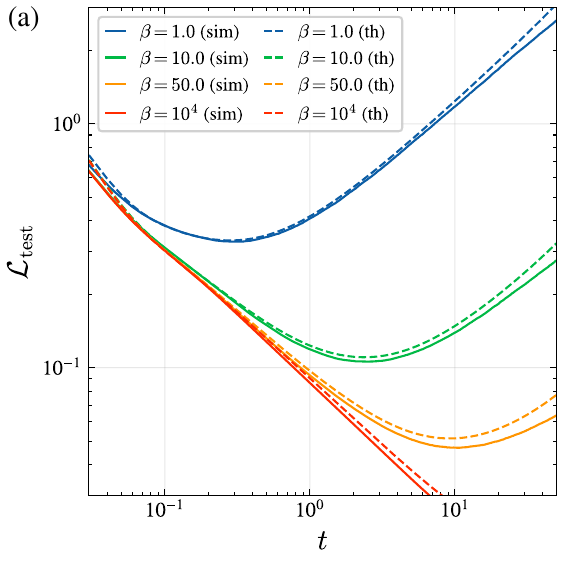}\label{fig:Test-error-as-function-of-beta}

}\hfill{}\subfloat{\includegraphics[width=0.95\columnwidth]{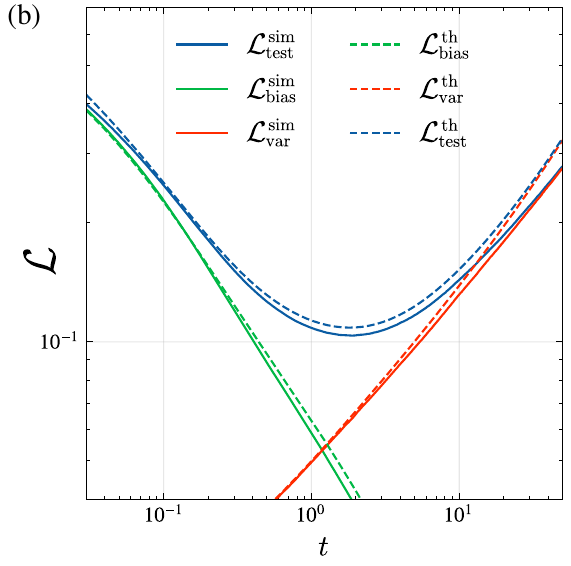}\label{fig:Test-error-decomposed}}\caption{(a) Test error for different strengths $2\beta^{-1}$ of the dynamic
noise with $\beta=10000$ (red), $\beta=50$ (green) and $\beta=10$
(blue). The solid curves show the theory \eqref{eq:test_error},
the dashed lines the simulation. The other parameters are $g\beta=10^{3},$$P=N=100$,
$\Lambda_{ij}=i^{-3/2}\delta_{ij}$. The time step used for the simulation
is $\protect\dd t=10^{-4}$, for the theory it is $\protect\dd t=10^{-2}.$
The disorder average in the simulation is taken over $10^{5}$ different
realizations of training data. (b) Bias-variance decomposition of
the test error. The blue curves show the full test error \eqref{eq:test_error}
(full curve simulation; dashed curve theory), the green curves show
the bias contribution to the test loss, the red show the variance
part (cf. \prettyref{eq:bias_var_decomp}). The expectation value
of the kernel follows the power law $\Lambda_{ij}=i^{-3/2}\delta_{ij}$.
The time step used for the simulation is $\protect\dd t=10^{-4},$for
the theory it is $\protect\dd t=10^{-2}.$ Disorder average in simulation
taken over $10^{5}$ different realizations of training data sets.
Other parameters $\beta=10$ and $g\beta=10^{3}$, system at the interpolation
threshold ($P=N=10^{2}$). }
\end{figure*}

The generalization error \eqref{eq:E_test} can be determined very
naturally in our theory as it can be expressed in terms of one of
the order parameters as
\begin{align}
\mathcal{L}_{\mathrm{test}} & =\frac{1}{2}\,\Bigg\langle\sum_{i=1}^{N}\psi_{i}\,\Big\langle\left(\bar{w}_{i}-w_{i}\right){}^{2}\Big\rangle_{\zeta}\,\Bigg\rangle_{\psi\sim\mathcal{N}(0,\eta_{i}\delta_{ij})}\label{eq:test_error}\\
 & =\frac{1}{2}\sum_{i=1}^{N}\eta_{i}\,\langle v_{i}(t)^{2}\rangle_{\zeta}=\frac{1}{2}\bar{C}(t,t)\quad,\nonumber 
\end{align}
allowing us to compute it together with \eqref{eq:variance_part}.
\prettyref{fig:Test-error-as-function-of-beta} shows the time evolution
of the test error as a function of time for different temperatures
$\beta^{-1}$. As the noise variance $2\beta^{-1}$ increases, the
generalization worsens as expected. For small times, the error declines,
but eventually the error increases again after reaching a local minimum.
To explain this behavior we decompose the test error 
\begin{equation}
\mathcal{L}_{\text{\text{test}}}=\frac{1}{2}\sum_{i=1}^{N}\eta_{i}\Bigg[\underbrace{\langle v_{i}(t)\rangle^{2}}_{\text{bias}}+\underbrace{\langle\delta v^{2}(t)\rangle}_{\text{variance}}\Bigg]\,,\label{eq:bias_var_decomp}
\end{equation}
into the bias and the variance contribution. As seen in \prettyref{fig:Test-error-decomposed},
the bias term decays in a monotonic fashion, independent of $\beta$,
while the fluctuations grow with higher temperatures and, so does
the variance term. In \prettyref{fig:Test-error-decomposed} we see
the minimum formed by this temporal bias-variance trade-off. While
initially the error is dominated by the decrease of the large modes'
discrepancies between student and teacher weights, the variance adds
up over time, eventually becoming the dominant contribution. This
result explains the practical use of early stopping, i.e. aborting
the training before equilibrium is reached, to be beneficial in the
presence of significant noise. As this observation is obtained with
weak regularization, it raises the question of whether this behavior
is altered, if the weight decay is regularized more strongly.

\begin{figure*}
\includegraphics[width=1\textwidth]{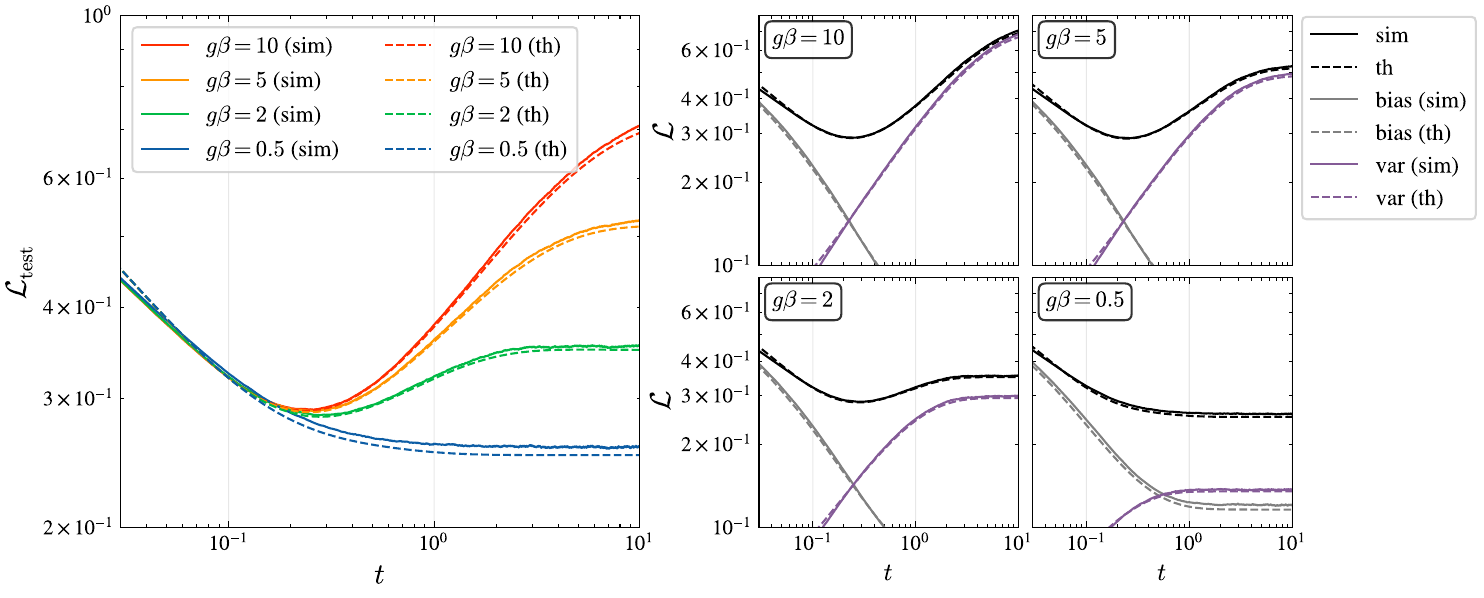}

\caption{Effect of regularization on early stopping. Left panel: Test error
for different strengths $g\beta$ of the regularization. The dashed
curves show the theory, the solid curves the simulation. Right panel:
Individual test error curves with their respective bias-variance decomposition.
Dashed curves are theory, solid curves simulation. Other parameters
for all panels $\beta=10^{0},$ $P=N=100$, $\Lambda_{ij}=i^{-3/2}\delta_{ij}$.
The time step used for the simulation was $\protect\dd t=10^{-4}$,
for the theory it is $\protect\dd t=10^{-2}.$ Disorder average in
simulation taken over $10^{5}$ different realizations of training
data sets.\protect\label{fig:Test-error-as-function-of-gbeta}}
\end{figure*}

The L2-regularization term $1/(g\beta)$ in the loss function \eqref{eq:def_H}
favors small weights and ensures the convergence of the weights even
for singular feature-feature matrices \prettyref{eq:Q_definition}.
It also prevents the weights from accumulating noise, acting as a
cutoff for the variance part of the generalization error. \prettyref{fig:Test-error-as-function-of-gbeta}
shows the test error for high but fixed noise and varying regularization.
Larger regularization $1/(g\beta)$ prevents the generalization error
from diverging. Interestingly, while this affects larger times, it
has small effect on the location and the magnitude of the minimum
loss. In a heuristic fashion this can again be explained with the
bias-variance trade-off: As seen in \prettyref{fig:Time-evolution-of-discrepancy}
the decay of the bias term is mainly governed by the fast modes, corresponding
to the large eigenvalues that adapt the strongest to data. As the
regularization counteracts the $P\eta K*$-term in the equation of
motion \eqref{eq:effective_eom} it affects those weights stronger
that have a small eigenvalue $\eta$; on short time-scales the bias
term is only weakly affected. The variance, on the other hand, is
a collective phenomenon. While the fast modes again contribute on
shorter timescales, because of the spectral bias, the slow modes become
more important over time because of their long range Green's functions
in \eqref{eq:variance_part} which counteract the small eigenvalues.
Their decay is, however, enforced by a larger regularization as can
be seen in \eqref{eq:green_fourier} and therefore has a restraining
effect on the variance part. This, in turn, limits the generalization
error on larger timescales. While the generalization is expected to
get worse if the regularization is too high, this result suggests
that regularization can be particularly helpful for training in the
presence of noise. 

In the noiseless case our setup is identical to the one \cite{spectral_canatar_2021}
consider in the stationary case and it can be shown that in the limit
$t\to\infty$, \prettyref{eq:test_error} recovers their results (see
\prettyref{sec:Stationary-Limit}). 

\section{Neural Scaling Laws}

\begin{figure}[!h]
\includegraphics[width=1\columnwidth]{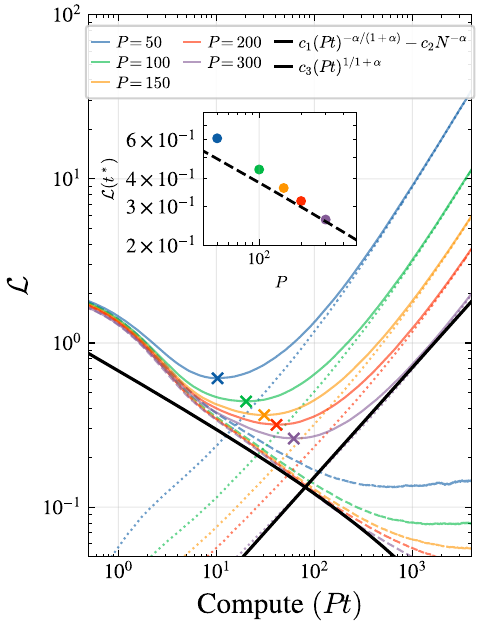}\caption{Neural scaling law for the test error as a function of training time
arises from superposition of exponentially decaying modes \eqref{eq:superposition_modes}.
Solid colored curves are the test error while the colored dashed and
dotted curves display the bias and variance error, respectively, for
different sample sizes $P$ obtained from numerical experiments. The
black descending curve is the limiting case for the bias part \eqref{eq:power_law_bias},
the ascending black curve the variance approximation \eqref{eq:power_law_variance}.
The other parameters are $\beta=10^{0},$ $g\beta=10^{5}$, $N=1000$,
$\Lambda_{ij}=i^{-3/2}\delta_{ij}$, $\protect\dd t=0.005$. Disorder
average in simulation taken over $10^{2}$ different realizations
of training data sets.\protect\label{fig:Power-laws} The inset plot
shows the minimal loss as a function of $P$. The colored dots correspond
to the crosses in the main plot, the dashed line is the power-law
approximation given in \eqref{eq:L_test_of_P_beta}.}
\end{figure}

We utilize the previously derived results to analyze the thermodynamic
limit of the model, which is known to exhibit power-laws. As the
full generalization error in general is a superposition of multiple
power-laws, it is beneficial to look at each component individually.
A detailed derivation of the power-laws given below can be found in
\prettyref{sec:Derivation-of-Scaling}.

\subsection{Bias error}

The bias error is bounded by the limiting Ornstein-Uhlenbeck process,
which is obtained from \eqref{eq:response_function} by approximating
$K(t-s)\simeq\delta(t-s)$. This approximation corresponds to the
neglect of fluctuations of $Q$, replacing it by $\langle Q\rangle$
in \eqref{eq:eom_w_main}. Since fluctuation corrections slow down
the learning dynamics, the bias will thus decay slower than 
\begin{align}
\mathcal{L}_{\mathrm{bias,OUP}}(t) & =\frac{1}{2}\sum\eta_{i}\bar{w}_{i}^{2}e^{-2\left(P\eta_{i}+\frac{1}{g\beta}\right)t}\label{eq:superposition_modes}
\end{align}
which in the $P,N\gg1$ limit decays as (see \prettyref{sec:Derivation-of-Scaling})
\begin{equation}
\mathcal{L}_{\mathrm{bias,OUP}}\sim c_{1}\,(Pt)^{-\alpha/(1+\alpha)}-c_{2}\,N^{-\alpha}\label{eq:power_law_bias}
\end{equation}
for $c_{1}$ and $c_{2}$ being positive constants. For constant $N$
this corresponds to the well known pareto frontier as function of
compute $f=6PNt$ as shown in \prettyref{fig:Power-laws}. For $P,t\to\infty$,
this recovers the result from \cite{bordelonDynamicalModelNeural2024a}
that $\mathcal{L}\sim N^{-\alpha}.$

\subsection{Variance error}

To investigate the behavior of the system in the presence of large
noise it is natural to look at the variance error in the limit $\beta\to0$.
In this regime, $\mathcal{L}_{\mathrm{var}}$ can be approximated
as
\begin{align*}
\mathcal{L}_{\mathrm{var}}(t) & \approx\mathcal{L}_{\beta}(t)\\
 & :=\frac{1}{\beta}\:\sum_{i=1}^{N}\eta_{i}\int_{0}^{t}\dd s\,G_{i}(t-s)^{2}\quad.
\end{align*}
Taking into account the same considerations as for $\mathcal{L}_{\mathrm{bias}}$
we find that 
\begin{equation}
\mathcal{L}_{\beta}\sim c_{3}\beta^{-1}\,P{}^{-\alpha/(1+\alpha)}\,t{}^{1/(1+\alpha)}\quad,\label{eq:power_law_variance}
\end{equation}
with $c_{3}=(1+\alpha)c_{1}.$This is in accordance with our intuition
that the variance decreases with  the amount of training data and
that the noise accumulates and hence grows over time. Interestingly,
 we find the same power-law as a function of $P$ as for the bias.

\prettyref{fig:Power-laws} shows the time evolution of $\mathcal{L}$
as a function of $Pt$ for different values of $P$. While the theoretical
predicted power-laws agree nicely, the $N^{-\alpha}$ correction cannot
be neglected and for finite system sizes the bias part has to be viewed
as a superposition of power-laws. 

\subsection{Early stopping}

\begin{figure}
\centering
\includegraphics[width=0.6\columnwidth]{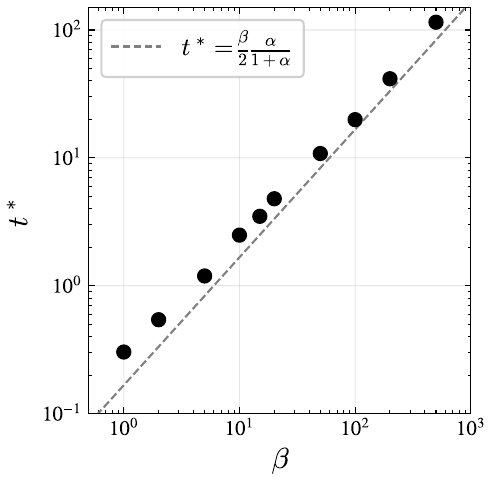}

\caption{Optimal stopping time as function of $\beta$. The points represent
the simulation, the dashed gray line the theoretical estimate based
on \eqref{eq:optimal_stopping_time}. Other parameters: $P=N=100$,
$\Lambda_{ij}=i^{-3/2}\delta_{ij}$, $g\beta=10^{3}$. The time step
used for the simulation was $\protect\dd t=10^{-4}$. Disorder average
in simulation taken over $5\times10^{3}$ different realizations of
training data sets.\protect\label{fig:t_opt-as-function-of-beta}}
\end{figure}

The optimal stopping time $t^{*}$ is given by the minimum of the
test error through
\[
0=\frac{\dd\left(\mathcal{L}_{\mathrm{bias}}(t)+\mathcal{L}_{\mathrm{var}}(t)\right)}{\dd t}\Bigg|_{t=t^{*}}\quad.
\]
Evaluating this equation, we find 
\begin{equation}
t^{*}=\frac{\beta}{2}\frac{\alpha}{1+\alpha}\label{eq:optimal_stopping_time}
\end{equation}
as well as for the generalization error evaluated at that point 
\begin{equation}
\mathcal{L}_{\mathrm{test}}(t^{*})\sim d_{1}\left(P\beta\right)^{-\alpha/(1+\alpha)}-d_{2}N^{-\alpha}\,.\label{eq:L_test_of_P_beta}
\end{equation}
We observe that the optimal stopping time $t^{*}$ is independent
of both $P$and $N$ but is only a function of $\beta$, a result
that can be verified in numerical simulations (see \prettyref{fig:t_opt-as-function-of-beta}).
This simple linear relationship \prettyref{eq:optimal_stopping_time}
should be considered in conjunction with \prettyref{eq:L_test_of_P_beta},
and may be interpreted as follows: under high-noise conditions, learning
must be terminated early to prevent the accumulation of noise, since
further training cannot recover additional structure from the data.
We further see in \prettyref{eq:L_test_of_P_beta} that the amount
of training data acts in a conjugate way to the noise $\beta$. We
see in \prettyref{fig:Power-laws} that this relationship holds approximately
in simulations, however finite size effects play an important role
in correctly estimating the optimal generalization error $\mathcal{L}_{\mathrm{test}}(t^{*})$. 

\section{Discussion}

In this work we present a self-consistent dynamical mean-field theory
to describe the training dynamics of kernel regression. The framework
is able to cover both gradient flow as well as Langevin stochastic
gradient descent and, at equilibrium, makes the link to Bayesian inference
of learning. Here we obtain an expression for the dynamics of the
generalization error that is applicable to the relevant case of power-law
distributed kernel spectra, as they are frequently encountered in
real data and in the kernel regime of trained networks \cite{Lee18,neural_jacot_2018}.
We demonstrate that this setting shows the phenomenon of early stopping,
where the test-loss is minimized before training has converged. This
work hence effectively extends on prior research such as \cite{generalization_krogh_1992}
and \cite{advaniHighdimensionalDynamicsGeneralization2020}, who
studied the case of learning dynamics on Gaussian i.i.d. data and
\cite{spectral_canatar_2021}, who studied the case of kernel regression
after convergence and whose results can be obtained from our theory.

The theory shows that the problem of learning is self-averaging, implying
that the data-averaged behavior is close to the behavior of a system
trained on a single data set. This allows us to average the dynamical
partition function over data disorder and the noise of Langevin training.
A technical advantage of the dynamical approach is that it avoids
the use of replicas \cite{dedominicisDynamicsSubstituteReplicas1978}.
We find that in the limit of large dimension $N\to\infty$, the full
statistics are entirely described by two self-averaging order parameters,
the trace of the system’s response function and the trace of the auto-correlation
of the parameters. The latter, moreover, is proportional to the test
error. This simplicity is due to averaging over the kernel directly,
while previous work relied on additional auxiliary fields \cite{bordelonDynamicalModelNeural2024a}.
Our work thus yields a compact and interpretable description of the
dynamics while still maintaining the full generality of the problem.
In the large $N$ limit, the central limit theorem guarantees the
Gaussianity of the process, which renders the performed variational
Gaussian approximation exact (see also \prettyref{app:VGA}). The
fact, that the test error appears in a natural way in the theory makes
it a particularly useful description to investigate the generalization
performance.

From the Gaussian action we derive an effective stochastic equation
of motion that exposes the spectral dependence of the mean evolution
of each mode on its eigenvalue. It yields a dynamical view on the
spectral bias of learning \cite{spectral_canatar_2021}, showing that
larger eigenmodes are not only learned with higher accuracy, but are
also learned earlier than smaller ones. In addition, we show that
the effective equations of motion of all modes become mutually statistically
independent. The relaxation time of each mode is slowed down by a
collective temporal non-local coupling that is caused by the disorder
average and whose temporal shape depends on the dynamics of all remaining
modes. This coupling can be identified as a geometric series of the
system’s perturbation response in frequency domain and acts like an
Onsager reaction term for the system \cite{Fischer91c}. Even though
modes become statistically independent, they are coupled indirectly
through this memory kernel: thus, modes that are learned more slowly,
indirectly also slow down the learning process of all remaining modes.

The joint dataset formally acts as a collective disorder which in
addition produces an effective noise term in the equation of motion.
This term, too, mediates an indirect coupling among modes, as the
mean-discrepancy of each mode contributes additively to this noise.
This temporally correlated and time-dependent noise, in turn, increases
the variance of the individual weights. As the effective noise is
linked to the generalization error, we find the full dynamics of the
bias-variance decomposition of the test error. It shows that the optimal
duration for learning that yields the minimal test error (early stopping)
results from an interplay between the effective noise, controlling
the gradual growth of the variance term of the loss, and the decay
time of the response function, which governs the decline of the bias
term of the generalization error. This early stopping time, in the
thermodynamic limit, becomes a constant that only depends on the decay
exponent of the kernel's eigenmodes and the inverse temperature. It
can be used to derive an expression for the optimal generalization
error as a superposition of competing scaling laws stemming from the
bias and variance terms respectively. 

Our setup can easily be extended to study the influence of other sources
of noise an example being static label noise as discussed in \prettyref{sec:label-noise}.
While it is restricted to linearized settings, such as kernel ridge
regression or linear regression, further work is needed to include
a combination of the dynamical formalism we developed here with contemporary
work on feature learning in non-linear networks \cite{Li21_031059,separation_seroussi_2023,Pacelli23_1497,Rubin25_arxiv,lauditiAdaptiveKernelPredictors2025}
to describe the learning dynamics of networks closer to real-life
settings and a possible extension is discussed for the case of committee
machines in \prettyref{sec:Extension-to-committee}. Recent work,
however, shows that the kernel limit of networks may indeed capture
the universal behavior in the large data limit \cite{Coppola_2026}.

\section*{Impact Statement}

This paper presents work whose goal is to advance the field of machine
learning. The work in particular aims to develop the theoretical understanding
of learning on high-dimensional data, with ubiquitous properties.
A fundamental understanding of these properties is helpful to provide
guarantees for learning and to guide future improvements of algorithms
and architectures. 

\section*{Acknowledgements}

This work was partly funded by the Deutsche Forschungsgemeinschaft
(DFG, German Research Foundation) - 368482240/GRK2416, the Helmholtz
Association Initiative and Networking Fund under project number SO-092
(Advanced Computing Architectures, ACA), the Deutsche Forschungsgemeinschaft
(DFG, German Research Foundation) as part of the SPP 2205 -- 533396241,
and the DFG grant 561027837/HE 9032/4-1. Open access publication funded
by the Deutsche Forschungsgemeinschaft (DFG, German Research Foundation)
-- 491111487. The authors gratefully acknowledge the computing time
granted by the JARA Vergabegremium and provided on the JARA Partition
part of the supercomputer JURECA at Forschungszentrum Jülich (computation
grant JINB33).

\bibliographystyle{icml2026}
\bibliography{statphys_bibliography,brain}
\newpage{}

\appendix
\onecolumn

\part*{Appendix}

\section{Generating functional in MSRDJ formalism \protect\label{sec:Generating-functional-in}}

Starting from the equation of motion in \ref{eq:of_motion_single}
we want to investigate the distribution on $v_{i}(t)$, to which purpose
we utilize the MSRDJ formalism \cite{helias20_970} to state a functional
$Z$ which generates the functional moments of the stochastic differential
equation
\begin{align}
Z & =\int\mathcal{D}v\mathcal{D}\tilde{v}\,\exp\big(S(v,\tilde{v})\big)\\
S(v,\tilde{v}) & =\int dt\,\sum_{i=1}^{N}\tilde{v}_{i}(t)\,\Big(\partial_{t}v_{i}(t)+\sum_{j=1}^{N}\Omega_{ij}v_{j}(t)-\frac{\overline{w}_{i}}{g\beta}\Big)+\frac{1}{\beta}\sum_{i=1}^{N}\tilde{v}_{i}(t)^{2}\,.\label{eq:action_def}
\end{align}
with \foreignlanguage{english}{$\Omega_{ij}:=Q_{ij}+\delta_{ij}/(g\beta)$
}. As we are ultimately interested in the disorder averaged learning
statistics on $v$ we rather need to consider $\langle Z\rangle$
and its derivatives. Considering the fact that, due to the Gaussianity
of the features $\psi$, the feature matrix $Q$ is a Wishart matrix,
we can write the disorder-averaged moment generating functional as
\begin{align}
\langle Z\rangle_{Q} & =\int\mathcal{D}v\mathcal{D}\tilde{v}\left\langle \exp(S(v,\tilde{v}))\right\rangle _{Q}\\
 & =\int\mathcal{D}v\mathcal{D}\tilde{v}\,\exp\big(\bar{S}(v,\tilde{v})\big)
\end{align}
with
\begin{equation}
\bar{S}(v,\tilde{v})=\int dt\,\sum\tilde{v}_{i}(t)\Big(\big[\partial_{t}+\frac{1}{g\beta}\big]\,v_{i}(t)-\frac{\overline{w}_{i}}{g\beta}+\frac{1}{\beta}\sum_{i=1}^{N}\tilde{v}_{i}(t)^{2}\,\Big)-\frac{P}{2}\,\ln\det\left(\mathbb{I}-\Lambda\left[\int dt\,\tilde{v}(t)v(t)^{\top}+v(t)\tilde{v}(t)^{\top}\right]\right)\label{eq:action_appendix_after_disorder}
\end{equation}
with $\mathbb{I}_{ij}=\delta_{ij}$. The last line is the cumulant-generating
function of the Wishart distribution; it can be obtained by a trivial
Gaussian integral over the Gaussian distributed modes $\psi$. We
here introduced the diagonal matrix $\Lambda_{ij}=\delta_{ij}\eta_{i}$.

This action is analytically intractable. However, we will be able
to treat this term analytically in the large $N$ limit by first introducing
a number of order parameters and subsequently performing a variational
Gaussian approximation (VGA), which will approximate the process $v$
by a Gaussian with non-zero mean. We will present this approach in
the following two sections.

\section{Identification of order parameters\protect\label{app:Identification-of-order-papameters}}

\textcolor{black}{The term after disorder in \eqref{eq:action_appendix_after_disorder},
rewritten as $-\frac{P}{2}\tr\,\ln\left(\mathbb{I}-\Lambda\left[\int dt\,\tilde{v}(t)v(t)^{\top}+v(t)\tilde{v}(t)^{\top}\right]\right)$
and expanded into the power series $\ln(1-x)=-\sum_{n=1}^{\infty}\,\frac{x^{n}}{n}$
yields with the short hand $J=\int\tv(t)v(t)^{\T}\,dt$}

\textcolor{black}{
\begin{align}
 & -\frac{1}{2}\ln\,\det\Big(\bI-\Lambda\,(J+J^{\T})\Big)\label{eq:disordered_term}\\
= & \frac{1}{2}\,\sum_{n=1}^{\infty}\,\frac{1}{n}\,\big[\Lambda\,(J+J^{\T})\big]^{n}\label{eq:binomial_like}\\
= & \frac{1}{2}\,\sum_{i}\,\eta_{i}\,J_{ii}+\sum_{i}\,\eta_{i}\,J_{ii}^{\T}\nonumber \\
 & +\frac{1}{4}\,\sum_{i}\,\sum_{j}\,\eta_{i}J_{ij}\,\eta_{j}J_{ji}+\frac{1}{4}\,\sum_{i}\,\sum_{j}\,\eta_{i}J_{ij}\,\eta_{j}J_{ji}^{\T}+\frac{1}{4}\,\sum_{i}\,\sum_{j}\,\eta_{i}J_{ij}^{\T}\,\eta_{j}J_{ji}+\frac{1}{4}\,\sum_{i}\,\sum_{i}\,\eta_{i}J_{ij}^{\T}\,\eta_{j}J_{ji}^{\T}\nonumber \\
 & +\ldots\,.\nonumber 
\end{align}
In this product the order of $J$ and $J^{\T}$ determines which self-averaging
terms arise, namely one has
\begin{align}
\big[\Lambda J\,\Lambda J\big]_{ij} & \equiv\iint\,\eta_{i}\tv_{i}(t)\underbrace{\sum_{k}\eta_{k}v_{k}(t)\,\tv_{k}(s)}_{=:R(t,s)}v_{j}(s)\,dt\,ds\,,\label{eq:second_order}\\
\big[\Lambda J^{\T}\Lambda J^{\T}\big]_{ij} & \equiv\iint\,\eta_{i}v_{i}(t)\underbrace{\sum_{k}\eta_{k}\tv_{k}(t)\,v_{k}(s)}_{=R(s,t)}\tv_{j}(s)\,dt\,ds\,,\nonumber \\
\big[\Lambda J^{\T}\Lambda J\big]_{ij} & \equiv\iint\,\eta_{i}v_{i}(t)\underbrace{\sum_{k}\eta_{k}\tilde{v}_{k}(t)\,\tv_{k}(s)}_{=:\tC(t,s)}v_{j}(s)\,dt\,ds\,,\nonumber \\
\big[\Lambda J\,\Lambda J^{\T}\big]_{ij} & \equiv\iint\,\eta_{i}\tv_{i}(t)\underbrace{\sum_{k}\eta_{k}v_{k}(t)\,v_{k}(s)}_{=:C(t,s)}\tv_{i}(s)\,dt\,ds\,.\nonumber 
\end{align}
So we obtain the rules
\begin{align*}
\Lambda J\Lambda J & \to\eta\,\tv\,R\,v\,,\\
\Lambda J^{\T}\Lambda J^{\T} & \to\eta\,v\,R^{\T}\,\tv\,,\\
\Lambda J^{\T}\Lambda J & \to\eta\,v\,\tC\,v,\\
\Lambda J\Lambda J^{\T} & \to\eta\,\tv\,C\,\tv.
\end{align*}
One notices that the remaining terms in \eqref{eq:second_order},
if written under the trace and after replacing the indicated parts
by the self-averaging quantity, mediate a coupling only between modes
of the same index, so they lead to an action that is diagonal.}

\textcolor{black}{}

\textcolor{black}{In the limit $N\to\infty$, we will find that $C$
and $R$ concentrate to their mean value. To demonstrate this, one
needs to show that the statistics of $(\tv_{k},v_{k})$ becomes independent
across $k$, as we will show in the following sections. If the distribution
of the $(\tv_{k},v_{k})$ factorizes across $k$, the concentration
of $C$ and $R$ follows from the central limit theorem. One additional
point to note here is that the power law $\eta_{k}=k^{-(1+\alpha)}$
is such that 
\begin{align}
\sum_{k=1}^{N}\eta_{k} & \simeq\int_{1}^{N}\,k^{-(1+\alpha)}\,dk=-\alpha^{-1}\,k^{-\alpha}\big|_{1}^{N}\stackrel{N\to\infty}{\to}\alpha^{-1}\label{eq:majorization}
\end{align}
converges to a constant for the assumed exponents $\alpha>0$. We
will find below that the correlation functions $\langle v_{k}v_{k}\rangle$
and response functions $\langle\tv_{k}v_{k}\rangle$ also decline
with $k$, which is due to the appearance of $\eta_{k}$ in the effective
equations of motion \eqref{eq:eom_delta_v_of_K}, so that \eqref{eq:majorization}
is a majorization of the definitions of $C$ and $R$ and thus both
also exist in the limit $N\to\infty$.}

\section{Variational Gaussian approximation\protect\label{app:VGA}}

\textcolor{black}{To treat the data-dependent term, one may perform
a variational Gaussian approximation (VGA) were modes are pairwise
independent, but may have mode-dependent variance and response functions
jointly contained in $G$ as well as a mode-dependent mean $(\bar{v},\bar{\tilde{v}})$.}

\textcolor{black}{This motivates the variational ansatz
\begin{align}
(v,\tv) & \sim\N\big[(\bar{v},\bar{\tilde{v}}),G\big]\propto\exp\left\{ -\frac{1}{2}\left(\begin{array}{c}
v-\bar{v}\\
\tv-\bar{\tilde{v}}
\end{array}\right)^{\top}G^{-1}\left(\begin{array}{c}
v-\bar{v}\\
\tv-\bar{\tilde{v}}
\end{array}\right)\right\} \label{eq:Gaussian_on_v_tilde_v}\\
G & =\left(\begin{array}{cc}
G^{(vv)} & G^{(v\tv)}\\
G^{(\tv v)} & G^{(\tv\tv)}
\end{array}\right)\,,\label{eq:G_matrix}
\end{align}
where each $G^{(xy)}$ is a diagonal $N\times N$ matrix, with elements
$G_{ij}^{(xy)}(t,s)=\delta_{ij}\,G_{i}^{(xy)}(t,s)$ for $x,y\in\{v,\tv\}$.
We will show a posteriori that this ansatz becomes exact in the limit
$N\to\infty$.}

\textcolor{black}{The parameters $\bar{v}$, $\bar{\tv}$, and $G$
will be determined from the equation of state, which arises from minimizing
the KL divergence between $p(v,\tv)\propto e^{S(v,\tv)}$ and the
Gaussian measure \eqref{eq:Gaussian_on_v_tilde_v}
\begin{align*}
\Gamma(\bar{v},\bar{\tilde{v}},G) & :=-\mathrm{KL}\left(\N\left[(\bar{v},\bar{\tilde{v}}),G\right]\Bigg|\Bigg|e^{\bar{S}(v,\tv)}\right)\\
 & =\big\langle\bar{S}(v,\tv)\big\rangle_{(v,\tv)\sim\N\big[(\bar{v},\bar{\tilde{v}}),G\big]}+\frac{1}{2}\ln\det\big[G\big]\,,
\end{align*}
where the latter term is the entropy of the Gaussian. The parameters
$\bar{v}$, $\bar{\tilde{v}}$, and $G$ that minimize the KL divergence
are determined by the stationarity conditions, also known as equations
of state,
\begin{align}
0\stackrel{!}{=} & \frac{\delta}{\delta\{\bar{v},\bar{\tilde{v}}\}}\,\left[\big\langle\bar{S}(v,\tv)\big\rangle_{(v,\tv)\sim\N\big[(\bar{v},\bar{\tilde{v}}),G\big]}\right]\Bigg|_{G^{(\tv\tv)}=0,\bar{\tilde{v}}=0}\,,\label{eq:eq_of_state_mean}\\
0\stackrel{!}{=} & \frac{\delta}{\delta G^{(xy)}}\left[\big\langle\bar{S}(v,\tv)\big\rangle_{(v,\tv)\sim\N\big[(\bar{v},\bar{\tilde{v}}),G\big]}+\frac{1}{2}\ln\det\big[G\big]\right]\Bigg|_{G^{(\tv\tv)}=0,\bar{\tilde{v}}=0}\,,\label{eq:eq_of_state_cov}
\end{align}
where we need to assure that $\bar{\tv}=0$ as well as $G^{(\tv\tv)}\equiv0$.}

\textcolor{black}{For the equations of state, we also need the inverse
of the matrix \eqref{eq:G_matrix}, which, due to its diagonality
in $i$, can be inverted for each $i$ separately as
\begin{align}
\frac{\delta}{\delta G_{i}^{(xy)}(t,s)}\,\frac{1}{2}\ln\det\big[G\big]\,\Big|_{G^{(\tv\tv)}=0} & =\frac{1}{2}\big[G\big]_{i}^{-1}(t,s)\label{eq:G_inv}\\
\text{with }\quad\big[G\big]^{-1} & =\left(\begin{array}{cc}
0 & \big[G^{(\tv v)}\big]^{-1}\\
\big[G^{(v\tv)}\big]^{-1} & -\big[G^{(v\tv)}\big]^{-1}G^{(vv)}\big[G^{(\tv v)}\big]^{-1}
\end{array}\right)\,.\nonumber 
\end{align}
Applied to \eqref{eq:second_order}, extended to arbitrary orders
of $n$, this means we only need to take into account terms of the
form
\begin{align}
\langle\tr(\Lambda J)^{n}\rangle_{(v,\tv)\sim\N\big[(\bar{v},\bar{\tilde{v}}),G\big]} & \to\big[\bar{R}\big]^{n}\,,\label{eq:n_th_power_terms}\\
\langle\tr(\Lambda J^{\T})^{n}\rangle_{(v,\tv)\sim\N\big[(\bar{v},\bar{\tilde{v}}),G\big]} & \to\big[\bar{R}^{\T}\big]^{n}\,,\nonumber \\
\langle\tr(\Lambda J)^{n-k}\,(\Lambda J^{\T})^{k}\rangle_{(v,\tv)\sim\N\big[(\bar{v},\bar{\tilde{v}}),G\big]} & \stackrel{1\le k\le n-1}{\to}\big[\bar{R}\big]^{n-k-1}\,\bar{C}\,\big[\bar{R}^{\T}\big]^{k-1}\,\bar{\tC}\,,\nonumber 
\end{align}
where
\begin{align}
\bar{R}^{\T}(s,t)=\bar{R}(t,s) & =\sum_{i}\eta_{i}\,\big[\bar{v}_{i}(t)\bar{\tilde{v}}_{i}(s)+G_{i}^{(v\tv)}(t,s)\big]\,,\label{eq:R_as_G}\\
\bar{C}(t,s) & =\sum_{i}\eta_{i}\,\big[\bar{v}_{i}(t)\bar{v}_{i}(s)+G_{i}^{(vv)}(t,s)\big]\,,\nonumber \\
\bar{\tC}(t,s) & =\sum_{i}\eta_{i}\,\big[\bar{\tilde{v}}_{i}(t)\bar{\tilde{v}}_{i}(s)+G_{i}^{(\tv\tv)}(t,s)\big]\,.\nonumber 
\end{align}
We may study explicitly the terms that we have dropped, when writing
the replacement rules \eqref{eq:n_th_power_terms} to show that they
either vanish completely or are subleading in the $N\to\infty$ limit.
Consider the example of $n=2$
\begin{align}
\tr\,\big[\Lambda J\,\Lambda J\big] & \equiv\sum_{ik}\,\eta_{i}\eta_{k}\,\iint\tv_{i}(t)v_{k}(t)\,\tv_{k}(s)v_{i}(s)\,\dd t\,\dd s\,.\label{eq:pairing_example}
\end{align}
A priori, there are two qualitatively different contributions. It
is easiest to first consider them perturbatively and subsequently
perform an infinite resummation of all such contributions, which we
technically do with the VGA. The first perturbative correction is
of the form}

\textcolor{black}{
\begin{align*}
\sum_{ik}\,\eta_{i}\eta_{k}\,\int\tv_{i}(t)v_{k}(t)\,\dd t\,\int\langle\tv_{k}(s)v_{i}(s)\rangle\,\dd s\,,
\end{align*}
which in principle presents a coupling term between modes $(i,k)$,
mediated by the time-averaged response function $\int\langle\tv_{k}(s)v_{i}(s)\rangle\,\dd s$.
The latter response function, however, vanishes, because in the Ito-convention
employed here, equal time response functions vanish (see, e.g., \cite{Coolen01,helias20_970}).
The second perturbative contribution is
\begin{align*}
 & \sum_{i}\,\eta_{i}\,\iint\tv_{i}(t)\,\sum_{k}\eta_{k}\,\langle v_{k}(t)\,\tv_{k}(s)\rangle v_{i}(s)\,\dd t\,\dd s\,,
\end{align*}
which is non-zero, because for $t>s$ the response function $\langle v_{k}(t)\,\tv_{k}(s)\rangle\neq0$
in general. Importantly, this term only mediates a self-coupling of
mode $i$ ($v_{i})$ to itself ($\tv_{i}$), thus it will not cause
correlations between modes. This term is the one taken into account
by the first replacement rule \eqref{eq:n_th_power_terms}.}

\textcolor{black}{Analogous considerations hold for all higher powers
$n>2$, where we get products of fields $\tv_{i}(r)v_{j}(r)\,\tv_{j}(s)v_{k}(s)\ldots\tv_{l}(t)v_{i}(t)$:
Whenever two fields with different mode indices are paired, e.g. $\langle\tv_{i}(r)v_{k}(s)\rangle$
this implies $r<s$ for the contribution not to vanish. But contracting
all remaining fields in pairs one always also gets a pairing such
as $\langle v_{j}(r)\tv_{j}(s)\rangle$, which requires $r>s$, so
together it will lead to vanishing contribution.}

\textcolor{black}{An additional perturbative contribution comes from
pairing two fields $v$. As a contribution to \eqref{eq:pairing_example}
one has
\begin{align}
 & \sum_{ik}\,\eta_{i}\eta_{k}\,\iint\tv_{i}(t)\,\tv_{k}(s)\,\langle v_{k}(t)v_{i}(s)\rangle\,\dd t\,\dd s\,,\label{eq:cross_corr}
\end{align}
which, for any pair of modes $(i,k)$ contributes a correlated Gaussian
noise (due to the appearance of $\tv_{i}(t)\,\tv_{k}(s)$) in proportion
to the correlation $\langle v_{k}(t)v_{i}(s)\rangle$ itself. Note
first, that a consistent solution is therefore one where modes are
uncorrelated. Note second that the driving noise $\zeta_{i}$ is uncorrelated
across modes, so it does not drive these correlations, so that vanishing
correlations are indeed the state assumed. Moreover, a single term
in the sum $\sum_{k}$ contributes this correlation. We may compare
its magnitude to the magnitude that arises from
\begin{align}
\tr\big[\Lambda J\Lambda J^{\T}\big] & =\sum_{ik}\,\eta_{i}\eta_{k}\,\iint\tv_{i}(t)v_{k}(t)\,v_{k}(s)\,\tv_{i}(s)\,\dd t\,\dd s\,.\label{eq:sub_leading_corr}
\end{align}
Here the corresponding contribution is
\begin{align}
\tr\big[\Lambda J\Lambda J^{\T}\big] & =\sum_{i}\,\eta_{i}\iint\tv_{i}(t)\,\sum_{k}\eta_{k}\langle v_{k}(t)\,v_{k}(s)\rangle\,\tv_{i}(s)\,\dd t\,\dd s\,,\label{eq:leading_corr}
\end{align}
where the summed variance $\sum_{k}\eta_{k}\langle v_{k}(t)\,v_{k}(s)\rangle$
of all modes causes a contribution to the variance of mode $i$ (due
to the appearance of two $\tv_{i}$). The latter term is hence of
order $\order(N)$ compared to the contribution \eqref{eq:cross_corr};
this shows that cross correlations between modes are smaller by an
order $\order(N)$ compared to the variance terms, so that in the
limit $N\to\infty$ the latter dominate.}

\textcolor{black}{We can also drop all terms where more than two blocks
of $J$ and $J^{\T}$ appear, such as for example $\langle\tr\,J\,J^{\T}JJ^{\T}\rangle=C\,\tC\,C\,\tC=\order(\tC^{2})$,
which is a term that will cancel because one $\tC$ corresponds to
the appearance of a pair of fields $\tv$ $\tv$ (causing an effective
Gaussian noise), but the second $\tC$ would correspond to the appearance
of a second pair $\langle\tv\tv\rangle\equiv0$, which vanishes by
the Ito convention (see \cite{Coolen01,helias20_970}). An analogous
consideration holds for any order of perturbation theory and therefore
also for the resummation of all such terms. In the VGA, this property
is included because after differentiation by $\tC$ (in the equation
of state \eqref{eq:eq_of_state_cov}) one needs to set all remaining
$\tC$ to zero due to the Ito convention. In the same manner, we need
to drop all terms with two or more factors of $\bar{\tilde{v}}$,
because we need to set $\bar{\tilde{v}}\equiv\langle\tv\rangle\equiv0$
after differentiation, again by the Ito convention.}

\textcolor{black}{In summary, all leading order terms in the limit
$N\to\infty$ contain concentrating quantities $C$ or $R$, respectively.
Dropping the sub-leading terms (such as \eqref{eq:sub_leading_corr}
as compared to \eqref{eq:leading_corr}), the remaining action contains
at most pairs of fields $v$ and $\tv$ and hence describes a Gaussian
action. Moreover, the sub-leading terms are also the ones that would
couple different modes $i\neq k$. This means that the Gaussian action
becomes one in which modes become independent in the limit (conditioned
on the value of the concentrating fields $C$ and $R$). This together
shows that the VGA in the $N\to\infty$ limit indeed becomes exact.}

The terms in \eqref{eq:n_th_power_terms} are all cyclicly invariant
under the trace. This means that a term with $n$ factors in total
and all factors identical ($J$ or $J^{\T}$) has a combinatorial
factor of $1$ as it appears exactly once when multiplying out \eqref{eq:binomial_like}
\begin{align*}
\frac{1}{n}\,\tr\,\big[\Lambda(J+J^{\T})\big]^{n} & =\frac{1}{n}\,\tr\,\big[\Lambda(J+J^{\T})\big]\,\ldots\,\big[\Lambda(J+J^{\T})\big]\,.
\end{align*}
By the cyclic invariance of the trace, a term that has $n-k$ factors
of one sort ($J$ or $J^{\T}$) and $k$ factors of the respective
other, appears $n$ times when multiplying out, because the position
where the change from one factor to the other takes place can appear
at $n$ positions; the respective change back is then fixed, because
$k$ is fixed. In addition, each term with $n$ factors $J$ or $J^{\T}$
comes with a factor $1/n$. 

So in total, the disorder term, in expectation under the Gaussian,
takes the form
\begin{align}
 & -\frac{1}{2}\,\Big\langle\ln\,\det\Big(\bI-\Lambda(J+J^{\T})\Big)\Big\rangle_{(v,\tv)\sim\N((\bar{v},\bar{\tilde{v}}),G)}\label{eq:exp_ln_det}\\
= & \iint\,\bar{R}(t,s)\,dt\,ds\nonumber \\
+ & \frac{1}{2}\,\iiint\,\bar{R}(t,u)\,\bar{R}(u,s)\,dt\,du\,ds+\frac{1}{2}\,\iiint\,\bar{C}(t,u)\,\bar{\tC}(u,s)\,dt\,du\,ds\nonumber \\
+ & \sum_{n=3}^{\infty}\,\frac{1}{n}\,\int\,\big[\bar{R}*\big]^{n}(t,t)\,dt\nonumber \\
+ & \frac{1}{2}\sum_{n=3}^{\infty}\,\sum_{k=1}^{n-1}\int\,\big[\bar{R}*\big]^{n-k-1}\,\bar{C}\,\big[\bar{R}^{\T}*\big]^{k-1}\,\bar{\tC}(\circ,t)\,dt\,,\nonumber 
\end{align}
where we write $\big[\bar{R}*\big]_{ts}^{k}=\int\cdots\int\,\bar{R}(t,u_{1})\cdots\bar{R}(u_{k-1},s)\,du_{1}\ldots du_{k-1}$.

We may now resum the terms that only depend on the response function
as $\sum_{n=1}^{\infty}\,\frac{1}{n}\,\int\,\big[\bar{R}\ast\big]^{n}\,dt=-\int\,\ln\,\big(1-\bar{R}\big)\,dt$.
Likewise we may combine the terms that contain up to a single factor
$\tC$ to obtain

\begin{align}
 & -\frac{1}{2}\,\Big\langle\ln\,\det\Big(\bI-\Lambda(J+J^{\T})\Big)\Big\rangle_{(v,\tv)\sim\N\big[(\bar{v},\bar{\tilde{v}}),G\big]}\label{eq:exp_ln_det_resummed}\\
= & \sum_{n=1}^{\infty}\,\frac{1}{n}\,\int\,\big[\bar{R}*\big]^{n}(t,t)\,dt\nonumber \\
+ & \frac{1}{2}\sum_{n=2}^{\infty}\,\sum_{k=1}^{n-1}\int\,\big[\bar{R}*\big]^{n-k-1}\,\bar{C}\,\big[\bar{R}^{\T}*\big]^{k-1}\,\bar{\tC}(\circ,t)\,dt\,.\nonumber 
\end{align}

\subsection{Equation of motion for the mean}

Using \eqref{eq:eq_of_state_mean} for $\delta/\delta\bar{\tilde{v}}$
we get
\begin{align*}
0 & \stackrel{!}{=}\frac{\delta}{\delta\bar{\tilde{v}}}\left[\big\langle\bar{S}(v,\tv)\big\rangle_{(v,\tv)\sim\N\big[(\bar{v},\bar{\tilde{v}}),G\big]}\right]\Bigg|_{G^{(\tv\tv)}=0,\bar{\tilde{v}}=0}\,.
\end{align*}
To evaluate this expression, we first compute the expectation value
of the Gaussian part
\begin{align}
 & \langle S_{0}\rangle_{(v,\tv)\sim\N\big[(\bar{v},\bar{\tilde{v}}),G\big]}\label{eq:exp_S0}\\
 & =\sum_{i=1}^{N}\,\iint\,\delta(t-s)\,\left\{ \left[\partial_{t}+\frac{1}{g\beta}\right]\,\left[\bar{\tv}_{i}(s)\bar{v}_{i}(t)+G_{i}^{(v\tv)}(t,s)\right]+\beta^{-1}\,\left[\bar{\tv}_{i}(t)\bar{\tv}_{i}(t)+G_{i}^{(\tv\tv)}(t,s)\right]\right\} \dd t\,\dd s\nonumber \\
 & -\int\frac{\bar{w}_{i}}{g\beta}\bar{\tv}_{i}(t)\,dt\,.\nonumber 
\end{align}
The contribution to the equation of state \eqref{eq:eq_of_state_mean}
is hence
\begin{align*}
 & \frac{\delta}{\delta\bar{\tilde{v}}_{i}(t)}\,\langle S_{0}\rangle_{(v,\tv)\sim\N\big[(\bar{v},\bar{\tilde{v}}),G\big]}\\
= & \left[\partial_{t}+\frac{1}{g\beta}\right]\,\bar{v}_{i}(t)+\underbrace{G_{i}^{(v\tv)}(t,t)}_{=0}-\frac{\bar{w}_{i}}{g\beta}\,,
\end{align*}
where $G_{i}^{(v\tv)}(t,t)=0$ by the Ito convention. The interaction
term with \eqref{eq:exp_ln_det_resummed} yields
\begin{align*}
\frac{\delta}{\delta\bar{\tilde{v}}_{i}(t)}\left\{ -\frac{P}{2}\,\Big\langle\ln\,\det\Big(\bI-\Lambda(J+J^{\T})\Big)\Big\rangle_{(v,\tv)\sim\N\big[(\bar{v},\bar{\tilde{v}}),G\big]}\right\} \Bigg|_{G^{(\tv\tv)}=0,\bar{\tilde{v}}=0} & =P\eta_{i}\int\,\sum_{n=0}^{\infty}\,\big[\bar{R}\ast\big]^{n}(t,s)\,\bar{v}_{i}(s)\,ds\,,
\end{align*}
where the factors $\eta_{i}$ and $\bar{v}_{i}(s)$ come from the
inner derivative $\delta\bar{R}(s,t)/\delta\bar{\tilde{v}}(t)$ from
\eqref{eq:R_as_G} and all terms vanish where at least one factor
$\tilde{C}$ or $\bar{\tilde{v}}$ remains after differentiation.
So together we get the equation of motion for the mean
\begin{align}
0= & \left[\partial_{t}+\frac{1}{g\beta}\right]\,\bar{v}_{i}(t)-\frac{\bar{w}_{i}}{g\beta}+P\eta_{i}\int\,\sum_{n=0}^{\infty}\,\big[\bar{R}\ast\big]^{n}(t,s)\,\bar{v}_{i}(s)\,ds.\label{eq:eom_mean}
\end{align}

\subsection{Solution of the response function}

The Gaussian part $S_{0}$ of \eqref{eq:action_appendix_after_disorder}
in the equation of state takes the form \eqref{eq:exp_S0}. Because
in the equation of state we set $\tC=0$ ultimately, all terms containing
$\tC$ drop out for the equation of state \eqref{eq:action_appendix_after_disorder}
for $G$; likewise, all terms $\propto\bar{\tilde{v}}$ vanish. So
one has 
\begin{align*}
\frac{\delta\langle S_{0}\rangle_{(v,\tv)\sim\N(0,G)}}{\delta G_{i}^{(v\tv)}(t,s)}\Bigg|_{G^{(\tv\tv)}=0} & =\delta(t-s)\,\left[\partial_{t}+\frac{1}{g\beta}\right]\,.
\end{align*}
The interaction term with \eqref{eq:exp_ln_det_resummed} yields
\begin{align*}
\frac{\delta}{\delta G_{i}^{(v\tv)}(t,s)}\left\{ -\frac{P}{2}\,\Big\langle\ln\,\det\Big(\bI-\Lambda(J+J^{\T})\Big)\Big\rangle_{(v,\tv)\sim\N\big[(\bar{v},\bar{\tilde{v}}),G\big]}\right\} \Bigg|_{G^{(\tv\tv)}=0} & =P\eta_{i}\,\sum_{n=0}^{\infty}\,\big[\bar{R}\ast\big]^{n}(t,s)\,,
\end{align*}
where the factor $\eta_{i}$ comes from the inner derivative $\delta\bar{R}/\delta G_{i}^{(v\tv)}$
from \eqref{eq:R_as_G}. We also note that the term $\bar{w}$ drops
out because we differentiated by $G^{(v\tv)}$.

This yields the equation of state \eqref{eq:eq_of_state_mean}
\begin{align}
0\stackrel{!}{=} & \delta(t-s)\,\left[\partial_{t}+\frac{1}{g\beta}\right]+P\eta_{i}\sum_{n=0}^{\infty}\,\big[\bar{R}\ast\big]^{n}(t,s)+\big[G_{i}^{(\tv v)}\big]^{-1}(t,s)\,,\label{eq:eq_state_response}
\end{align}
which is written as an implicit equation as

\begin{align}
\left[\partial_{t}+\frac{1}{g\beta}\right]\,G_{i}^{(v\tv)}(t,s)+P\eta_{i}\int\,\sum_{n=0}^{\infty}\,\big[\bar{R}\ast\big]^{n}(t,u)\,G_{i}^{(v\tv)}(u,s)\,\dd u & =-\delta(t-s)\,.\label{eq:self_consistent_G}
\end{align}
Moving to Fourier domain and resumming the geometric series one has
\begin{align}
\left[i\omega+\frac{1}{g\beta}\right]\,G_{i}^{(v\tv)}(\omega)+P\eta_{i}\,\frac{1}{1-\bar{R}(\omega)}\,G_{i}^{(v\tv)}(\omega) & =-1\,.\label{eq:G_resummed_Fourier}
\end{align}
We notice that the self-feedback mediated by the response functions
is an alternating series, since $G_{i}^{(v\tv)}(t,s)<0$. The latter
equation needs to be solved self-consistently together with the definition
of the response function $\bar{R}$ by \eqref{eq:R_as_G} at the saddle
point $\bar{\tilde{v}}=0$
\begin{align*}
\bar{R}(t,s) & =\sum_{i=1}^{N}\eta_{i}G_{i}^{(v\tv)}(t,s)\,.
\end{align*}
Likewise, we formally get an equation of motion for $\bar{\tilde{v}}(t)$
by differentiating by $\bar{v}(t)$, which indeed admits the solution
$\bar{\tilde{v}}(t)=0$, since the resulting equation contains at
least one factor of $\bar{\tilde{v}}$ in each term.

\subsection{Solution of the correlation function}

The equation of state with regard to $G_{i}^{(\tv\tv)}$ yields the
self-consistency equation for the autocorrelation. The Gaussian part
yields \eqref{eq:exp_S0}
\begin{align*}
\frac{\delta\langle S_{0}\rangle_{(v,\tv)\sim\N\big[(\bar{v},\bar{\tilde{v}}),G\big]}}{\delta G_{i}^{(\tv\tv)}(t,s)}\Bigg|_{G^{(\tv\tv)}=0,\bar{\tilde{v}}=0} & =\beta^{-1}\,\delta(t-s)\,,
\end{align*}
and the term \eqref{eq:exp_ln_det_resummed} yields

\begin{align*}
 & \frac{\delta}{\delta G_{i}^{(\tv\tv)}(t,s)}\left\{ -\frac{P}{2}\,\Big\langle\ln\,\det\Big(\bI-\Lambda(J+J^{\T})\Big)\Big\rangle_{(v,\tv)\sim\N\big[(\bar{v},\bar{\tilde{v}}),G\big]}\right\} \Bigg|_{G^{(\tv\tv)}=0,\bar{\tilde{v}}=0}\\
 & =\frac{P}{2}\eta_{i}\,\sum_{n=2}^{\infty}\,\sum_{k=1}^{n-1}\Big\{\big[\bar{R}*\big]^{n-k-1}\,\bar{C}\,\big[\bar{R}^{\T}*\big]^{k-1}\Big\}(t,s)\,,
\end{align*}
so that together we have the equation of state \eqref{eq:eq_of_state_cov}
\begin{align}
0 & =\beta^{-1}\,\delta(t-s)+\frac{P}{2}\eta_{i}\sum_{n=2}^{\infty}\sum_{k=1}^{n-1}\left\{ \,\big[\bar{R}*\big]^{n-k-1}\,\bar{C}\,\big[\bar{R}^{\T}*\big]^{k-1}\right\} (t,s)-\frac{1}{2}\big[G^{(v\tv)}\big]^{-1}G^{(vv)}\big[G^{(\tv v)}\big]^{-1}\,.\label{eq:self-consistency_correlation}
\end{align}
We may insert \eqref{eq:eq_state_response} to get the equation of
motion for $G^{(vv)}$ as
\begin{align}
 & \iint\,\left[\left(\partial_{t}+\frac{1}{g\beta}\right)\delta(t-t^{\prime})+P\eta_{i}\,\sum_{n=0}^{\infty}\,\big[\bar{R}\ast\big]^{n}(t,t^{\prime})\right]\label{eq:eq_of_state_C}\\
 & \times\left[\left(\partial_{s}+\frac{1}{g\beta}\right)\delta(s-s^{\prime})+P\eta_{i}\,\sum_{n=0}^{\infty}\,\big[\bar{R}\ast\big]^{n}(s,s^{\prime})\right]\,G_{i}^{(vv)}(t^{\prime},s^{\prime})\,\dd t^{\prime}\,\dd s^{\prime}\nonumber \\
= & 2\beta^{-1}\,\delta(t-s)+P\eta\sum_{n=2}^{\infty}\,\sum_{k=1}^{n-1}\left\{ \,\big[\bar{R}*\big]^{n-k-1}\,\bar{C}\,\big[\bar{R}^{\T}*\big]^{k-1}\right\} (t,s)\,.\nonumber 
\end{align}

\section{Effective equation of motion\protect\label{sec:Effective-equation-of-motion}}

We may interpret the equation of motion for the response function
\eqref{eq:self_consistent_G} together with the equation of state
for the correlation function \eqref{eq:eq_of_state_C} as both resulting
from the same the effective stochastic equation of motion
\begin{align}
\Big[\partial_{t}+\frac{1}{g\beta}\Big]\,\delta v_{i}(t)+P\eta_{i}\int\,\sum_{n=0}^{\infty}\,\big[\bar{R}\ast\big]^{n}(t,s)\,\delta v_{i}(s)\,\dd s & =\xi_{i}(t)\,.\label{eq:of_motion}
\end{align}
The mean, in addition, follows the equation \eqref{eq:eom_mean},
which may be included here to yield the effective equation given in
the main text \eqref{eq:effective_eom}.

The resummed response function $K(t,s)=\sum_{n=0}^{\infty}\,\big[\bar{R}\ast\big]^{n}(t,s)$
may be regarded at the solution $K$ of the implicit equation
\begin{align}
K(t,s) & =\int\dd u\,\bar{R}(t-u)\,K(u-s)+\delta(t-s)\label{eq:effective_K}
\end{align}
so that one has
\begin{align}
\Big[\partial_{t}+\frac{1}{g\beta}\Big]\,\delta v_{i}(t)+P\eta_{i}\int\,K(t,s)\,\delta v_{i}(s)\,\dd s & =\xi_{i}(t)\,,\label{eq:eom_delta_v_of_K}
\end{align}
where the variance of the noise, according to the right hand side
of \eqref{eq:eq_of_state_C}, is given by
\begin{align}
\langle\xi_{i}(t)\xi_{j}(s)\rangle & =\delta_{ij}\,\Big[2\beta^{-1}\,\delta(t-s)+P\eta_{i}\sum_{n=2}^{\infty}\,\sum_{k=1}^{n-1}\left\{ \big[\bar{R}*\big]^{n-k-1}\,\bar{C}\,\big[\bar{R}^{\T}*\big]^{k-1}\right\} (t,s)\,\Big].\label{eq:corr_noise_eff}
\end{align}
The double sum may be rewritten as
\begin{align*}
 & \sum_{n=2}^{\infty}\,\sum_{k=1}^{n-1}\,\big[\ldots\big]^{n-k-1}\,\,\big[\ldots\big]^{k-1}\\
= & \sum_{n=0}^{\infty}\,\sum_{k=0}^{n}\,\big[\ldots\big]^{n-k}\,\,\big[\ldots\big]^{k}\\
= & \sum_{l=0}^{\infty}\,\sum_{k=0}^{\infty}\,\,\big[\ldots\big]^{l}\,\,\big[\ldots\big]^{k}\,,
\end{align*}
by renaming $n-k=:l$, so that we may write the correlator \eqref{eq:corr_noise_eff}
with \eqref{eq:effective_K} as
\begin{align}
\langle\xi_{i}(t)\xi_{j}(s)\rangle= & \delta_{ij}\left[2\beta^{-1}\,\delta(t-s)+P\eta_{i}\,\big\{ K\ast\bar{C}\ast K^{\T}\big\}(t,s)\right]\quad.\label{eq:xi_xi_correlator}
\end{align}

\section{Generalization Error}

\subsection{Variance part}

Writing the effective equation of motion of the discrepancy \ref{eq:of_motion}
as 
\[
L[\delta v]=\xi
\]
where $L$ is a linear operator and $\xi$ is an inhomogeneity, the
solution can be written in terms of the Green's function as
\[
\delta v_{i}(t)=\int\dd s\,G_{ij}(t-s)\xi_{j}(s)
\]
 with
\[
L[G]=\delta_{ij}\delta(t-s)\quad.
\]
The Green's function can be interpreted as the $\expval{v\tv}$-correlator
in \ref{eq:response_function} and the variance can be obtained by
taking the average utilizing \eqref{eq:corr_noise_eff}
\begin{align*}
\expval{\delta v_{i}(t)\delta v_{j}(s)} & =\iint\dd{u}\dd{u'}G_{ik}(t-u)\expval{\xi_{k}(u)\xi_{l}(u')}G_{jl}(s-u')\\
 & =2\beta^{-1}\int_{0}^{\min(t,s)}\dd{u}G_{i}(t-u)G_{j}(u-s)\delta_{ij}\\
 & +P\eta_{i}\int_{0}^{t}\int_{0}^{s}\dd{u}\dd{u'}G_{i}(t-u)\left\{ K\ast\bar{C}\ast K^{\top}\right\} (u,u')G_{j}(s-u')\delta_{ij}\quad.
\end{align*}

\subsection{Bias part}

As for the variance part, the mean $\bar{v}$ obeys the same differential
equation as $\delta v$
\[
L[\bar{v}]=\frac{\bar{w}}{g\beta}\quad.
\]
Hence, the solution takes on the form 
\begin{align*}
\bar{v} & =\bar{v}_{\text{hom}}+\bar{v}_{\text{part}}\\
 & =\bar{v}_{\text{hom}}+\frac{\bar{w}}{g\beta}\int_{0}^{t}\dd s\,G(t-s)\quad.
\end{align*}

While for $\delta v$ the homogeneous solution that satisfies the
boundary condition is $\delta v\equiv0$ by design, the boundary conditions
for $\bar{v}$ cannot be satisfied with a vanishing homogeneous solution,
if the system starts in a \textit{tabula rasa} state, i.e. $\bar{v}(0)=\bar{w}$.
In the stationary state the homogeneous parts of the fields have to
vanish as before, i.e. $\bar{v}_{\text{hom}}(\infty)=0$. As the Green's
function satisfies the homogeneous equation, the full solution is
found to be
\[
\bar{v}_{i}(t)=\bar{w}_{i}G_{i}(t)+\frac{\bar{w_{i}}}{g\beta}\int_{0}^{t}\dd s\:G(t-s)\quad.
\]

\subsection{Generalization error}

The generalization error is given with \eqref{eq:test_error} as 
\begin{align*}
\mathcal{L}_{\text{test}}(t) & =\frac{1}{2}\bar{C}(t,t)\\
 & =\frac{1}{2}\sum_{i}\eta_{i}\expval{\delta v_{i}(t)^{2}}+\frac{1}{2}\sum_{i}\eta_{i}\bar{v}_{i}(t)^{2}\quad.
\end{align*}

\section{Bayesian interpretation of the partition function\protect\label{sec:Bayesian-interpretation-of}}

Writing the thermodynamic partition function as 
\begin{align*}
\mathcal{Z} & =\int\dd w\:e^{-\beta H(w,\mathcal{D})}\\
 & =\int\dd w\:e^{-\frac{\beta}{2}\,\sum_{\alpha=1}^{P}(y_{\alpha}-f_{\alpha})^{2}-\frac{1}{2g}\sum_{i=1}^{N}w_{i}^{2}}
\end{align*}
with the Hamiltonian $H$ defined in \eqref{eq:def_H} allows for
a probabilistic viewpoint of the partition function as 
\[
\mathcal{Z}\propto\int\dd w\:\mathcal{N}(y|f(w),\beta^{-1})\,\mathcal{N}(w|0,g)\,.
\]
Because $f$ is a linear function of $w$ its distribution is again
Gaussian with variance
\[
\expval{f(x_{\mu})f(x_{\nu})}_{p(w)}=\expval{w^{\top}\psi(x_{\mu})\psi(x_{\nu})^{\top}w}_{p(w)}=g\sum_{i}\psi_{i}(x_{\mu})\psi_{i}(x_{\nu})\,,
\]
where the latter can be identified as the kernel $\kappa(x_{\mu},x_{\nu})$
\ref{eq:kernel_kappa}. From here it follows
\[
p(y)=\mathcal{N}(y|0,\kappa+\beta^{-1}\mathbb{\mathbb{I}})\,,
\]
where the Langevin noise $\beta^{-1}$ acts as a regulator to the
distribution of the $y$ akin to a static label noise. 

\section{Derivation of Scaling Laws\protect\label{sec:Derivation-of-Scaling}}

We obtain a natural lower bound on the generalization error by considering
the limit in which we neglect the disorder of the matrix $Q$ in \prettyref{eq:eom_w_main},
replacing $Q\to\langle Q\rangle$, so that the mutual slowing down
between individual nodes vanishes. Effectively, this treats the weights
as independent Ornstein-Uhlenbeck like processes. In this limit we
have 
\[
K(t-s)=\delta(t-s)
\]
and 
\[
G_{i}(t-s)=\Theta(t-s)e^{-\left(P\eta_{i}+\frac{1}{g\beta}\right)(t-s)}
\]
with $\Theta$ the Heaviside function following the convention $\Theta(0)=0$. 

\subsection{Bias part\protect\label{subsec:Bias-part-scaling-law}}

The bias part $\mathcal{L}_{\mathrm{bias}}$ of the loss simplifies
to ($t\gg1$) 
\[
\mathcal{L}_{\mathrm{bias}}(t)=\frac{1}{2}\sum_{i}\eta_{i}\bar{w}_{i}^{2}e^{-2\left(P\eta_{i}+\frac{1}{g\beta}\right)t}\quad.
\]
A power-law arises in the case where regularization is small, i.e.
$g\beta P\eta_{i}\gg1$. After averaging over $\bar{w}_{i}\sim\mathcal{N}(0,1)$,
in the large $N$ limit, the sum can be treated as an integral
\begin{align*}
\expval{\mathcal{L}_{\mathrm{bias}}(t)}_{\bar{w}} & \approx\frac{1}{2}\sum_{i=1}^{N}\eta_{i}e^{-2P\eta_{i}t}\\
 & =\frac{1}{2}\sum_{i=1}^{N}i^{-1-\alpha}e^{-2Pi^{-1-\alpha}t}\\
 & \approx\frac{1}{2}\int_{1}^{N}i^{-1-\alpha}e^{-2Pi^{-1-\alpha}t}\dd i\\
 & =\frac{1}{2}\frac{(2Pt)^{-\alpha/(1+\alpha)}}{1+\alpha}\int_{2Pt/N^{1+\alpha}}^{2Pt}u^{-1/(1+\alpha)}e^{-u}\,\dd u\\
 & \overset{Pt\gg1}{=}\frac{1}{2}\frac{(2Pt)^{-\alpha/(1+\alpha)}}{1+\alpha}\left(\Gamma\left(\frac{\alpha}{1+\alpha}\right)-\gamma\left(\frac{\alpha}{1+\alpha},2Pt/N^{1+\alpha}\right)\right)\\
 & =\frac{1}{2}\frac{(2Pt)^{-\alpha/(1+\alpha)}}{1+\alpha}\Gamma\left(\frac{\alpha}{1+\alpha}\right)-\frac{1}{2}N^{-\alpha}/\alpha+\order\left(N^{-\alpha}(Pt)\right)\,,
\end{align*}
where $\gamma(a,b)$ is the lower incomplete Gamma function \cite{DLMF12}.
Note that this only works for $\alpha>0$. This limit yields a similar
result as eq. 125 in \cite{bordelonDynamicalModelNeural2024a}.

\subsection{Variance part\protect\label{subsec:Variance-part-scaling-law}}

The variance part of the generalization error $\mathcal{L}_{\mathrm{var}}$
also shows power-law behavior under certain assumptions. If we want
to investigate early stopping in the regime of high temperature, the
main contribution comes from the $\beta$-dependent part of $\mathcal{L}_{\mathrm{var}}$
\begin{align*}
\mathcal{L}_{\mathrm{var}}(t) & \approx\mathcal{L}_{\beta}(t):=\frac{1}{\beta}\:\sum_{i=1}^{N}\eta_{i}\int_{0}^{t}\dd s\,G_{i}(t-s)^{2}\\
 & \approx\frac{1}{\beta}\:\sum_{i=1}^{N}\eta_{i}\int_{0}^{t}\dd se^{-2P\eta_{i}(t-s)}\\
 & =\frac{1}{2P\beta}\sum_{i=1}^{N}\left[1-e^{-2P\eta_{i}t}\right]
\end{align*}
In the same manner as before, this can rewritten as
\begin{align*}
2P\beta\mathcal{L}_{\beta}(t) & =N-\sum_{i=1}^{N}e^{-2P\eta_{i}t}\\
 & \approx N-\int_{1}^{N}e^{-2Pi^{-1-\alpha}t}\dd i\\
 & =N-\frac{(2Pt)^{1/(1+\alpha)}}{1+\alpha}\int_{2Pt/N^{1+\alpha}}^{2Pt}u^{-1/(1+\alpha)-1}e^{-u}\,\dd u\\
 & =N-\frac{(2Pt)^{1/(1+\alpha)}}{1+\alpha}\left(\Gamma\left(-\frac{1}{1+\alpha}\right)-\gamma\left(-\frac{1}{1+\alpha},\frac{2Pt}{N^{1+\alpha}}\right)\right)\\
 & \approx N-\frac{(2Pt)^{1/(1+\alpha)}}{1+\alpha}\Gamma\left(-\frac{1}{1+\alpha}\right)+N+\order\left(N^{-\alpha}(Pt)\right)\\
 & =(2Pt)^{1/(1+\alpha)}\Gamma\left(\frac{\alpha}{1+\alpha}\right)+\order\left(N^{-\alpha}(Pt)\right)\quad.
\end{align*}

\subsection{Optimal generalization error\protect\label{subsec:Optimal-generalization-error}}

The optimal stopping point can be estimated as the stationary point
of the test error that satisfies 
\begin{align*}
0 & =\frac{\dd\left(\mathcal{L}_{\mathrm{bias}}(t)+\mathcal{L}_{\beta}(t)\right)}{\dd t}\Bigg|_{t=t^{*}}\\
 & =\frac{\dd}{\dd t}\left(\frac{(2Pt)^{-\alpha/(1+\alpha)}}{1+\alpha}+\frac{(2Pt)^{1/(1+\alpha)}}{P\beta}\right)\Bigg|_{t=t^{*}}\\
 & =\frac{\dd}{\dd t}\left(\frac{(2P)^{1/(1+\alpha)}t^{-\alpha/(1+\alpha)}}{2P(1+\alpha)}+\frac{(2Pt)^{1/(1+\alpha)}}{P\beta}\right)\Bigg|_{t=t^{*}}\\
 & =-\frac{\alpha(t^{*})^{-\alpha/(1+\alpha)-1}}{2(1+\alpha)^{2}}+\frac{1}{\beta}\frac{(t^{*})^{-\alpha/(1+\alpha)}}{1+\alpha}\\
\Leftrightarrow t^{*} & =\frac{\beta}{2}\frac{\alpha}{(1+\alpha)}\quad.
\end{align*}
Evaluating $\mathcal{L}_{\,\mathrm{test}}$ at this point yields
\begin{align*}
\mathcal{L}_{\,\mathrm{test}}(t^{*}) & \approx\Gamma\left(\frac{\alpha}{1+\alpha}\right)\left(\frac{(2Pt^{*})^{1/(1+\alpha)}}{2P\beta}+\frac{1}{2}\frac{(2Pt^{*})^{-\alpha/(1+\alpha)}}{1+\alpha}\right)-\frac{1}{2\alpha}N^{-\alpha}\\
 & =\frac{1}{2}(P\beta)^{-\alpha/(1+\alpha)}\Gamma\left(\frac{\alpha}{1+\alpha}\right)\left(\left(\frac{\alpha}{(1+\alpha)}\right)^{1/(1+\alpha)}+\frac{\left(\frac{\alpha}{(1+\alpha)}\right)^{-\alpha/(1+\alpha)}}{1+\alpha}\right)-\frac{1}{2\alpha}N^{-\alpha}\\
 & =\frac{1}{2}\Gamma\left(\frac{\alpha}{1+\alpha}\right)\left(\frac{P\beta\alpha}{(1+\alpha)}\right)^{-\alpha/(1+\alpha)}-\frac{1}{2\alpha}N^{-\alpha}\\
 & =\frac{1}{2}\Gamma\left(\frac{\alpha}{1+\alpha}\right)\left(2Pt^{*}\right)^{-\alpha/(1+\alpha)}-\frac{1}{2\alpha}N^{-\alpha}
\end{align*}
We first note, that this approximation can work when the spectrum
decays faster than $1/i$ which is in agreement with the fact that
for a slower decay $\mathcal{L}_{\mathrm{bias}}$ and $\mathcal{L}_{\beta}$
would become extensive in the number of parameters. The optimal stopping
time is the later the faster the spectrum decays corresponding to
the system being able to improve generalization for longer during
training and minimizing $\mathcal{L}_{\mathrm{test}}$ at a lower
value. It is however limited by the inverse temperature which induces
an upper threshold for $t^{*}$ and thereby an implicit timescale
on the system as the time where the bias has decayed such that the
accumulated fluctuations are of the same order of magnitude. Surprisingly,
the optimal stopping point is independent of the number of parameters
as well as the number of samples to this order of approximation.

On the level of the generalization error we recover the intuitions
that more samples as well as more parameters aid generalization. The
sample size $P$ appears as a conjugate to the inverse temperature
$\beta$ in the way that in this regime noise and sample scale exactly
anti-proportional.

Our results agree with \cite{bordelonDynamicalModelNeural2024a} in
the case that $t$ and $N$ are large with the corresponding other
parameters go to infinity but differ in the case of finite $P$ while
$t,N\to\infty$ as we are not able to take the $t\to\infty$ limit
easily. This is due to the fact, that we consider the $\beta$-dependent
part as the dominating contribution while it is not present in their
work and hence the other contributions to the loss become dominant.
In the noiseless case we can take the stationary limit easily and
recover results from the literature as discussed in \prettyref{sec:Stationary-Limit}.

\section{Disorder average with label noise\protect\label{sec:label-noise}}

To include a read-out noise on the labels \eqref{eq:label_def},
i.e. 
\[
y^{\mu}=\bar{w}^{\T}\psi(x^{\mu})+\epsilon^{\mu}
\]
with $\epsilon^{\mu}\sim\mathcal{N}(0,\sigma^{2})$ the Loss function
$H$ \eqref{eq:def_H} takes the form

\begin{align*}
H:= & \frac{1}{2}\sum_{\mu=1}^{P}\left[\sum_{i=1}^{N}(\overline{w}_{i}-w_{i})\psi_{i}(x_{\mu})-\epsilon_{\mu}\right]^{2}+\frac{1}{2g\beta}\sum_{i=1}^{N}w_{i}^{2}\\
= & \frac{1}{2}\sum_{i,j=1}^{N}(\overline{w}_{i}-w_{i})\sum_{\mu}\psi_{i}(x_{\mu})\psi_{j}(x_{\mu})\,(\overline{w}_{j}-w_{j})\\
 & -\sum_{i=1}^{N}(\overline{w}_{i}-w_{i})\sum_{\mu=1}^{P}\,\psi_{i}(x_{\mu})\epsilon_{\mu}+\frac{1}{2g\beta}\sum_{i=1}^{N}w_{i}^{2}+\mathrm{const.}
\end{align*}
which leads to the equation of motion 
\begin{align*}
\frac{\partial}{\partial t}w_{i}(t) & =\sum_{j=1}^{N}\sum_{\mu=1}^{P}\psi_{i}(x_{\mu})\psi_{j}(x_{\mu})\,(\overline{w}_{j}-w_{j})-\frac{1}{g\beta}w_{i}+\sum_{\mu=1}^{P}\,\psi_{i}(x_{\mu})\epsilon_{\mu}+\zeta_{i}(t)\,.
\end{align*}
So the new action \eqref{eq:action_def} becomes with $v=\bar{w}-w$
\begin{align*}
S(v,\tilde{v}) & =\int\dd t\sum_{i=1}^{N}\tilde{v}_{i}(t)\left(\partial_{t}v_{i}(t)+\sum_{j=1}^{N}\sum_{\mu=1}^{P}\psi_{i}(x_{\mu})\psi_{j}(x_{\mu})\,v_{j}(t)+\frac{v_{i}-\overline{w}_{i}}{g\beta}+\sum_{\mu=1}^{P}\,\psi_{i}(x_{\mu})\epsilon_{\mu}\right)+\frac{1}{\beta}\sum_{i=1}^{N}\tilde{v}_{i}(t)^{2}\,.
\end{align*}
The term $\sum_{\mu=1}^{P}\,\psi_{i}(x_{\mu})\epsilon_{\mu}$ is linear
in $\epsilon$, therefore, we can average the moment generating function
$Z$ directly over the label noise and obtain for the action
\begin{align*}
S(v,\tilde{v}) & =\int\dd t\sum_{i=1}^{N}\tilde{v}_{i}(t)\left(\partial_{t}v_{i}(t)+\sum_{j=1}^{N}\sum_{\mu=1}^{P}\psi_{i}(x_{\mu})\psi_{j}(x_{\mu})\,v_{j}(t)+\frac{v_{i}-\overline{w}_{i}}{g\beta}\right)+\frac{1}{\beta}\sum_{i=1}^{N}\tilde{v}_{i}(t)^{2}\\
 & +\frac{\sigma^{2}}{2}\,\sum_{i=1}^{N}\iint\tilde{v}_{i}(t)\,\tilde{v}_{j}(s)\,\dd t\,\dd s\,\sum_{\mu=1}^{P}\,\psi_{i}(x_{\mu})\psi_{j}(x_{\mu})\,.
\end{align*}
Structurally, this appears as an additive factor to the Langevin noise
of the OUP, i.e.
\[
\frac{2}{\beta}\delta_{ij}\delta(t-s)\to\frac{2}{\beta}\delta_{ij}\delta(t-s)+\sigma^{2}\sum_{\mu=1}^{P}\,\psi_{i}(x_{\mu})\psi_{j}(x_{\mu})
\]
introducing a frozen noise term coupling both parameters and time
points. Taking the disorder average over the $\psi$ akin to the derivation
of \eqref{sec:Generating-functional-in}, we get
\begin{align}
\bar{S}(v,\tilde{v}) & =\int\dd t\sum\tilde{v}_{i}(t)\,\left(\left[\partial_{t}+\frac{1}{g\beta}\right]\,v_{i}(t)-\frac{1}{g\beta}\overline{w}_{i}\right)+\frac{1}{\beta}\sum_{i=1}^{N}\tilde{v}_{i}(t)^{2}\nonumber \\
 & -\frac{P}{2}\ln\det\left(\mathbb{I}-\Lambda\,\left[\int\tilde{v}(t)v(t)^{\top}+v(t)\tilde{v}(t)^{\top}dt+\sigma^{2}\iint\tilde{v}(t)\,\tilde{v}(s)^{\top}\,\dd t\,\dd s\,\right]\right)\,.\label{eq:S_disorder_averaged}
\end{align}

\subsection{Identification of order parameters}

Similar to \eqref{app:Identification-of-order-papameters} we introduce
$J:=\int\tilde{v}(t)v(t)^{\top}\,\dd t$ and $M:=\sigma^{2}\iint\tilde{v}(t)\,\tilde{v}(s)^{\top}\,\dd t\,\dd s$.
Note that each factor $J$ comes with a single time integral, while
each factor $M$ comes with two time integrals. We expand the $\ln\det$
with help of the logarithm series as

\begin{align}
 & -\frac{1}{2}\ln\,\det\Big(\bI-\Lambda\,(J+J^{\T}+M)\Big)\label{eq:disordered_term-1}\\
= & \frac{1}{2}\,\sum_{n=1}^{\infty}\,\frac{1}{n}\,\tr\,\big[\Lambda\,(J+J^{\T}+M)\big]^{n}\label{eq:binomial_like-1}\\
= & \frac{1}{2}\,\sum_{i}\,\eta_{i}\,J_{ii}+\sum_{i}\,\eta_{i}\,J_{ii}^{\T}+\sum_{i}\,\eta_{i}\,M_{ii}\nonumber \\
 & +\frac{1}{4}\,\sum_{i}\,\sum_{j}\,\eta_{i}J_{ij}\,\eta_{j}J_{ji}+\frac{1}{4}\,\sum_{i}\,\sum_{j}\,\eta_{i}J_{ij}\,\eta_{j}J_{ji}^{\T}+\frac{1}{4}\,\sum_{i}\,\sum_{j}\,\eta_{i}J_{ij}^{\T}\,\eta_{j}J_{ji}+\frac{1}{4}\,\sum_{i}\,\sum_{i}\,\eta_{i}j_{ij}^{\T}\,\eta_{j}j_{ji}^{\T}\nonumber \\
 & +\frac{1}{4}\,\sum_{i}\,\sum_{j}\,\eta_{i}J_{ij}\,\eta_{j}M_{ji}+\frac{1}{4}\,\sum_{i}\,\sum_{j}\,\eta_{i}J_{ij}^{\T}\,\eta_{j}M_{ji}+\frac{1}{4}\,\sum_{i}\,\sum_{j}\,\eta_{i}M_{ij}\,\eta_{j}J_{ji}+\frac{1}{4}\,\sum_{i}\,\sum_{j}\,\eta_{i}M_{ij}\,\eta_{j}J_{ji}^{\T}\nonumber \\
 & +\frac{1}{4}\,\sum_{i}\,\sum_{j}\,\eta_{i}M_{ij}\,\eta_{j}M_{ji}\\
 & +\ldots\,.\nonumber 
\end{align}
So the following products arise\allowdisplaybreaks{

\begin{alignat}{1}
\big[\Lambda J\,\Lambda J\big]_{ij} & \equiv\iint\,\eta_{i}\tv_{i}(t)\underbrace{\sum_{k}\eta_{k}v_{k}(t)\,\tv_{k}(s)}_{=:R(t,s)}v_{j}(s)\,\dd t\,\dd s\,,\label{eq:second_order-1}\\
\big[\Lambda J^{\T}\Lambda J^{\T}\big]_{ij} & \equiv\iint\,\eta_{i}v_{i}(t)\underbrace{\sum_{k}\eta_{k}\tv_{k}(t)\,v_{k}(s)}_{=:R(s,t)}\tv_{j}(s)\,\dd t\,\dd s\,,\nonumber \\
\big[\Lambda J^{\T}\Lambda J\big]_{ij} & \equiv\iint\,\eta_{i}v_{i}(t)\underbrace{\sum_{k}\eta_{k}\tilde{v}_{k}(t)\,\tv_{k}(s)}_{=:\tC(t,s)}v_{j}(s)\,dt\,ds\,,\nonumber \\
\big[\Lambda J\,\Lambda J^{\T}\big]_{ij} & \equiv\iint\,\eta_{i}\tv_{i}(t)\underbrace{\sum_{k}\eta_{k}v_{k}(t)\,v_{k}(s)}_{=:C(t,s)}\tv_{j}(s)\,\dd t\,\dd s\,,\nonumber \\
\big[\Lambda J\,\Lambda M\big]_{ij} & \equiv\iiint\,\eta_{i}\tv_{i}(t)\underbrace{\sum_{k}\eta_{k}v_{k}(t)\,\tv_{k}(s)}_{=:R^{\T}(t,s)}\sigma^{2}\tv_{j}(u)\,\dd t\,\dd s\,\dd u,\nonumber \\
\big[\Lambda J^{\T}\,\Lambda M\big]_{ij} & \equiv\iiint\,\eta_{i}v_{i}(t)\underbrace{\sum_{k}\eta_{k}\tv_{k}(t)\,\tv_{k}(s)}_{=:\tC(t,s)}\sigma^{2}\,\tv_{j}(u)\,\dd t\,\dd s\,\dd u,\nonumber \\
\big[\Lambda M\,\Lambda J\big]_{ij} & \equiv\iiint\,\eta_{i}\tv_{i}(t)\sigma^{2}\underbrace{\sum_{k}\eta_{k}\tv_{k}(s)\,\tv_{k}(u)}_{=:\tC(s,u)}v_{j}(u)\,\dd t\,\dd s\,\dd u,\nonumber \\
\big[\Lambda M\,\Lambda J^{\T}\big]_{ij} & \equiv\iiint\,\eta_{i}\tv_{i}(t)\sigma^{2}\underbrace{\sum_{k}\eta_{k}\tv_{k}(s)\,v_{k}(u)}_{=:R(u,s)}\tv_{j}(u)\,\dd t\,\dd s\,\dd u,\nonumber \\
\big[\Lambda M\,\Lambda M\big]_{ij} & \equiv\iiiint\,\eta_{i}\tv_{i}(t)\sigma^{2}\underbrace{\sum_{k}\eta_{k}\tv_{k}(s)\,\tv_{k}(u)}_{=:\tC(s,u)}\sigma^{2}\tv_{j}(r)\,\dd t\,\dd s\,\dd u\,\dd r\quad.\nonumber 
\end{alignat}
}In the following, we leave it implicit that a factor $J$ comes
with a single time integral and a factor $M$ with two independent
time integrals. With this shorthand, we obtain the rules
\begin{align*}
J\Lambda J & \to\tv Rv,\\
J^{\T}\Lambda J^{\T} & \to vR^{\T}\tv\,,\\
J^{\T}\Lambda J & \to v\tC v\,,\\
J\Lambda J^{\T} & \to\tv C\tv\,,\\
J\Lambda M & \to\tv R^{\T}\tv\,,\\
J^{\T}\Lambda M & \to v\tC\tv\,,\\
M\Lambda J & \to\tv\tC v\,,\\
M\Lambda J^{\T} & \to\tv R\tv\\
M\Lambda M & \to\tv\tC\tv\,.
\end{align*}
We may group these rules according to which fields remain to better
see the structure
\begin{align*}
1. & \{JJ\,,MJ\} & \to & \tv\{R,\tC\}v\,,\\
2. & \{J^{\T}J^{\T}\,,J^{\T}M\} & \to & v\{R^{\T},\tC\}\tv\,,\\
3. & J^{\T}J\, & \to & v\tC v\,,\\
4. & \{JJ^{\T},JM,MJ^{\T},MM\} & \to & \tv\{C,R^{\T},R,\tC\}\tv\,.
\end{align*}
In the VGA we only need to keep those terms where at most a single
factor $\tC$ or $\tv$ appears, because in the equation of state
we will differentiate by either of these variables once and subsequently
set it to zero -- any term containing more than a single power of
$\tC$ or $\tv$ thus vanishes.

We may therefore
\begin{itemize}
\item drop terms with factors $MM$, because they yield one $\tC$ and two
remaining $\tv$, which would cause another $\tC$ ultimately
\item drop terms which include more than a single factor $J^{\T}J$
\item terms that include a single factor $J^{\T}J$ already contain a single
$\tC$, so they can only be combined from left with $J^{\T}J^{\T}$
and from right with $JJ$ (and not with $J^{\T}M$ or $MJ$, because
this would yield another factor $\tC$)
\item drop terms with more than a single occurrence of $MJ$ or $J^{\T}M$
\item drop terms with more than a single occurrence from set 4. $\{JJ^{\T}$,
$JM$, $MJ^{\T}$\}, because each such occurrence causes a pair of
$\tv$ to remain on the left and on the right, which ultimately needs
to be contracted to yield a $\tC$ on either side
\item terms which include one element of set 4. $\{JJ^{\T},JM,MJ^{\T}\}$
can only be combined with factors $JJ$ from left and $J^{\T}J^{\T}$
from the right from set 1. (and not with $MJ$ or $JM$), for otherwise
there would be more than a single $\tC$ in the end
\end{itemize}
The remaining terms are

\begin{align}
\langle\tr(\Lambda J)^{n}\rangle_{(v,\tv)\sim\N\big[(\bar{v},\bar{\tilde{v}}),G\big]} & \to\big[\bar{R}\big]^{n}\,,\label{eq:n_th_power_terms-1}\\
\langle\tr(\Lambda J^{\T})^{n}\rangle_{(v,\tv)\sim\N\big[(\bar{v},\bar{\tilde{v}}),G\big]} & \to\big[\bar{R}^{\T}\big]^{n}\,,\nonumber \\
\langle\tr(\Lambda J)^{n-k}\,(\Lambda J^{\T})^{k}\rangle_{(v,\tv)\sim\N\big[(\bar{v},\bar{\tilde{v}}),G\big]} & \stackrel{1\le k\le n-1}{\to}\big[\bar{R}\big]^{n-k-1}\,\bar{C}\,\big[\bar{R}^{\T}\big]^{k-1}\,\bar{\tC}\,,\nonumber \\
\langle\tr(\Lambda J^{\T})^{n-k}\,(\Lambda J)^{k}\rangle_{(v,\tv)\sim\N\big[(\bar{v},\bar{\tilde{v}}),G\big]} & \stackrel{1\le k\le n-1}{\to}\big[\bar{R}^{\T}\big]^{n-k-1}\,\bar{\tC}\,\big[\bar{R}\big]^{k-1}\,\bar{C}\,,\nonumber \\
\langle\tr(\Lambda J)^{n-k}\,M\,(\Lambda J^{\T})^{k-1}\rangle_{(v,\tv)\sim\N\big[(\bar{v},\bar{\tilde{v}}),G\big]} & \stackrel{1\le k\le n-1}{\to}\big[\bar{R}\big]^{n-k-1}\,\bar{R}\,\big[\bar{R}^{\T}\big]^{k-2}\,\bar{\tC}\,,\nonumber \\
\langle\tr(\Lambda J^{\T})^{n-k}\,M\,(\Lambda J^{\T})^{k-1}\rangle_{(v,\tv)\sim\N\big[(\bar{v},\bar{\tilde{v}}),G\big]} & \to\big[\bar{R}^{\T}\big]^{n-k-1}\,\bar{\tC}\,\big[\bar{R}^{\T}\big]^{k}\nonumber \\
\langle\tr(\Lambda J)^{n-k}\,M\,(\Lambda J)^{k-1}\rangle_{(v,\tv)\sim\N\big[(\bar{v},\bar{\tilde{v}}),G\big]} & \to\big[\bar{R}\big]^{n-k}\,\bar{\tC}\,\big[\bar{R}\big]^{k-1}\,
\end{align}
Due to the different time arguments in the $\sigma^{2}\iint\tv(t)\tv(s)\dd t\dd s$-term,
we get additional integrals compared to \eqref{eq:pairing_example}.
In particular for 
\begin{align*}
\langle\tr(JMJ)\rangle & =\sigma^{2}\idotsint\langle\tv(t)v(t)\tv(s)\tv(u)\tv(w)v(w)\rangle\,\dd t\,\dd s\,\dd u\,\dd w\\
 & =\sigma^{2}\idotsint\langle v(t)\tv(s)\rangle\langle\tv(u)\tv(w)\rangle\langle v(w)\tv(t)\rangle\,\dd t\,\dd s\,\dd u\,\dd w\\
 & \sim\sigma^{2}\idotsint R(t-s)\tC(u,w)R(w-t)\,\dd t\,\dd s\,\dd u\,\dd w
\end{align*}
with all $\Lambda$ to be imagined in the right places. Collecting
all higher orders and summing over them would yield a term (after
differentiation w.r.t. $\tC(u,w)$) 
\[
\sigma^{2}\int K^{*}(w-s)\,\dd s
\]
with $K^{*}\equiv\sum_{k=1}^{\infty}R^{k}=K-\delta$, where interestingly
this term is $u$ independent and integrated over $s$. Similarly
for the $\langle\tr(\Lambda J^{\T})^{n-1}M\rangle_{(v,\tv)\sim\N\big[(\bar{v},\bar{\tilde{v}}),G\big]}$
terms, one gets a factor of the form
\[
\sigma^{2}\int K^{*\T}(w-s)\,\dd s\,.
\]
The reason to introduce $K^{*}$ is that the zeroth order term ($\langle M\rangle=\sigma^{2}\tC$)
can only appear in one of the three terms which is why we will keep
it separate.

Finally, there are the mixed terms
\[
\langle\tr(\Lambda J)^{n-k}\,M\,(\Lambda J^{\T})^{k-1}\rangle_{(v,\tv)\sim\N\big[(\bar{v},\bar{\tilde{v}}),G\big]}\stackrel{1\le k\le n-1}{\to}\big[\bar{R}\big]^{n-k-1}\,\bar{R}\,\big[\bar{R}^{\T}\big]^{k-2}\,\bar{\tC}\,,
\]
 which produce (after differentiation w.r.t. $\tC(u,w)$) 
\begin{align*}
\sum_{n=2}^{\infty}\sum_{k=1}^{n-1}R^{n-k}(t,s)\left(R^{\T}\right)^{k}(u,w)\tC(t,w) & =\sum_{n=1}^{\infty}\sum_{k=1}^{n}R^{n-k+1}(t,s)\left(R^{\T}\right)^{k}(u,w)\tC(t,w)\\
 & =\sum_{l=1}^{\infty}\sum_{k=1}^{\infty}R^{l}(t,s)\left(R^{\T}\right)^{k}(u,w)\tC(t,w)\\
 & =\left(K^{*}\right)^{\T}(u-w)\tC(w,t)K^{*}(t-s)\quad.
\end{align*}
Putting everything together this yields:
\begin{align*}
\left\langle \xi_{i}(t)\xi_{j}(s)\right\rangle  & =\delta_{ij}\left[2\beta^{-1}\,\delta(t-s)+P\eta_{i}\,\left\{ K\ast\bar{C}\ast K^{\T}\right\} (t,s)\right]\\
 & +P\sigma^{2}\eta_{i}\delta_{ij}\left[1+\int K^{*}(t-u)\,\dd u+\int K^{*\top}(w-s)\,\dd w+\int K^{*}(t-u)\,\dd u\,\int K^{*\top}(w-s)\,\dd w\right]\\
 & =\delta_{ij}\left[2\beta^{-1}\,\delta(t-s)+P\eta_{i}\,\left\{ K\ast\bar{C}\ast K^{\T}\right\} (t,s)\right]\\
 & +P\sigma^{2}\eta_{i}\delta_{ij}\left[1+\int K^{*}(t-u)\dd u\right]\left[1+\int K^{*}(s-u)\dd u\right]^{\T}\\
 & =\delta_{ij}\left[2\beta^{-1}\,\delta(t-s)+P\eta_{i}\,\left\{ K\ast\bar{C}\ast K^{\T}\right\} (t,s)\right]\\
 & +P\sigma^{2}\eta_{i}\delta_{ij}\left[\int K(t-u)\dd u\right]\left[\int K(s-u)\dd u\right]^{\T}\quad.
\end{align*}
\begin{figure}
\centering
\includegraphics[width=0.5\textwidth]{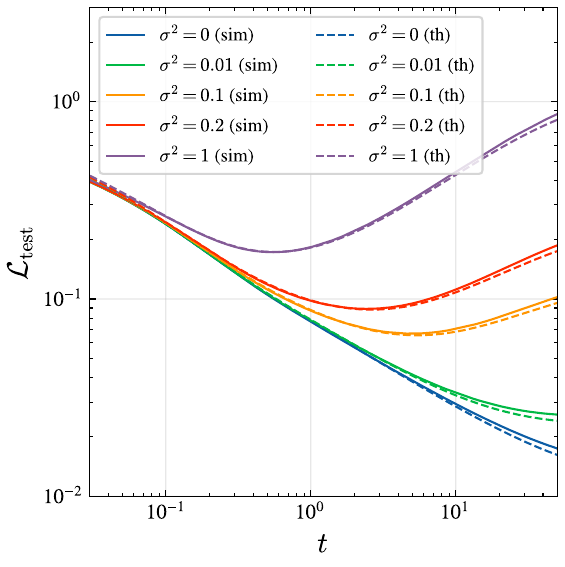}\caption{Test error for different strengths $\sigma^{2}$ of the label noise
with $\sigma^{2}=0.0$ (blue), $\sigma^{2}=0.01$ (green), $\sigma^{2}=0.1$
(orange), $\sigma^{2}=0.2$ (red) and $\sigma^{2}=1.0$ (purple).
The solid curves show the simulation, the dashed curves the theory.
The other parameters are $g\beta=10^{3},$$P=N=100$, $\Lambda_{ij}=i^{-3/2}\delta_{ij}$
and $\beta=10^{3}$. The time step used for the simulation is $\protect\dd t=10^{-4}$,
for the theory  $\protect\dd t=10^{-2}.$ The disorder average in
the simulation is taken over $10^{5}$ different realizations of training
data.}
\end{figure}

\section{Stationary Limit\protect\label{sec:Stationary-Limit}}

We want to show, that our approach is consistent with previous results
from the literature, where as a reference we show the equivalence
to eq. 4 of \cite{spectral_canatar_2021} who consider the static
case in the noiseless limit $\beta\to\infty$. First, we note that
$\frac{1}{g\beta\hat{K}(0)}$ can be identified with their $\kappa$
as they fulfill the same self-consistency equation, which can be seen
from rewritten for $\omega=0$ as 
\begin{align*}
\kappa:=\frac{1}{g\beta\hat{K}(0)} & =\frac{1}{g\beta}\left(1-\sum_{i}\eta_{i}\hat{G}_{i}(0)\right)\\
 & =\frac{1}{g\beta}\left(1+\sum_{i}\frac{\eta_{i}}{\frac{1}{g\beta}+P\eta_{i}\hat{K}_{i}(0)}\right)\\
 & =\frac{1}{g\beta}+\frac{1}{g\beta\hat{K}(0)}\sum_{i}\frac{\eta_{i}}{\frac{1}{g\beta\hat{K}(0)}+P\eta_{i}}\\
 & =\frac{1}{g\beta}+\kappa\,\sum_{i}\frac{\eta_{i}}{\kappa+P\eta_{i}}\,,
\end{align*}
where $1/g\beta$ is equivalent to their $\lambda$ and we used the
Fourier transform of \eqref{eq:K_of_R} using \eqref{eq:R_of_G}
in the first and \eqref{eq:green_fourier} in the second step. From
this equation we can see the useful relation 
\[
\hat{G}_{i}(0)\equiv g\beta\frac{\kappa}{\kappa+P\eta_{i}}\quad.
\]

First note that the test error in the stationary case also decomposes
into the bias and variance part 
\begin{align*}
\mathcal{L}_{\mathrm{test}}(t\to\infty)= & \frac{1}{2}\bar{C}(t=s\to\infty)\\
= & \frac{1}{2}\sum_{i}\eta_{i}\langle v_{i}(t\to\infty)\rangle^{2}+\frac{1}{2}\sum_{i}\eta_{i}\langle\delta v_{i}(t\to\infty)\rangle^{2}\,.
\end{align*}
The bias part in the long time limit can be evaluated using \eqref{eq:stationary_value}
by writing it as 
\begin{equation}
\mathcal{L}_{\mathrm{bias}}(t\to\infty)=\frac{1}{2}\sum_{i}\eta_{i}\frac{\hat{G}_{i}^{2}(0)}{\left(g\beta\right)^{2}}\bar{w}_{i}^{2}\equiv\frac{1}{2}\kappa^{2}\sum_{i}\eta_{i}\left(\frac{\bar{w}_{i}}{\kappa+P\eta_{i}}\right)^{2}\,,\label{eq:L_bias_stationary}
\end{equation}
in accordance to supplementary eq. (76) of \cite{spectral_canatar_2021}.

To evaluate the variance part, we use that for $\langle\delta v(t)\delta v(s)\rangle$
becoming stationary in both $t$ and $s$, we have from \eqref{eq:variance_part}
\begin{align*}
\frac{1}{2}\langle\delta v_{i}(t\to\infty)\delta v_{i}(s\to\infty)\rangle & =P\eta_{i}\hat{G}_{i}(0)\,\hat{K}(0)\,\mathcal{L}_{\mathrm{test}}(t\to\infty)\,\hat{K}^{\top}(0)\,\hat{G}_{i}^{\top}(0)\,,
\end{align*}
so that we get

\begin{align}
\mathcal{L}_{\mathrm{test}}(t\to\infty) & =\mathcal{L}_{\mathrm{bias}}(t\to\infty)+P\,\sum_{i}\eta_{i}^{2}\hat{G}_{i}^{2}(0)\hat{K}^{2}(0)\mathcal{L}_{\mathrm{test}}(t\to\infty)\label{eq:L_test_error_stationary}\\
 & =:\mathcal{L}_{\mathrm{bias}}(t\to\infty)+\gamma\,\mathcal{L}_{\mathrm{test}}(t\to\infty)\quad,\nonumber 
\end{align}
where we introduced a new parameter $\gamma$ in the same manner as
in \cite{spectral_canatar_2021}. This can be seen by looking at 
\[
P\,\sum_{i}\eta_{i}^{2}\hat{G}_{i}^{2}(0)\hat{K}^{2}(0)=P\sum_{i}\eta_{i}^{2}\left(g\beta\hat{K}(0)\right)^{2}\left(\frac{\kappa}{\kappa+P\eta_{i}}\right)^{2}=\sum_{i}\frac{P\eta_{i}^{2}}{\left(\kappa+P\eta_{i}\right)^{2}}
\]
which is the $\gamma$ defined there.

Plugging \eqref{eq:L_bias_stationary} into \eqref{eq:L_test_error_stationary}
and solving for $\mathcal{L}_{\mathrm{test}}(t\to\infty)$ we get
\[
\mathcal{L}_{\mathrm{test},\infty}=\frac{1}{1-\gamma}\sum_{i}\frac{\kappa^{2}\eta_{i}\bar{w}_{i}^{2}}{\left(\kappa+P\eta_{i}\right)^{2}}\,,
\]
which agrees to the result from \cite{spectral_canatar_2021}.

\section{Extension to committee machines\protect\label{sec:Extension-to-committee}}

The framework derived can be extended also to non-linear settings
and we want to introduce a possible extension here. Consider the following
teacher-student setup \cite{Krogh91} with 
\begin{align}
f_{\mu}:=f(w,x_{\mu})= & \phi(w^{\T}x_{\mu})\,,\\
y_{\mu}:= & \phi(\bar{w}^{\T}x_{\mu})\,,
\end{align}
with data $x_{\mu}$ and $\mu$ indexes the training points. The labels
$y_{\mu}$ are generated by the teacher weights $\bar{w}\in\mathbb{R}^{d}$
and $w\in\mathbb{R}^{d}$ are the student weights to be learned. The
train set for this regression task is given by $\cD=\{(x_{\mu},y_{\mu})\}_{\mu=1,\ldots,P}$
and $(x_{*},y_{*})\notin\mathcal{D}$ is a test point. We assume $x_{\mu i},x_{\ast i}\stackrel{\text{i.i.d.}}{\sim}\N(0,d^{-1})$
to be drawn from the same distribution. We solve the regression task
on the teacher-student setup with a squared loss
\begin{equation}
H(w,\mathcal{D}):=\frac{1}{2}\,\sum_{\mu=1}^{P}\,\big(y_{\mu}-f(w,x_{\mu})\big)^{2}+\frac{1}{2g\beta}\,\|w\|^{2}.\label{eq:def_H-1}
\end{equation}
Training the system with stochastic Langevin gradient descent yields
\begin{align*}
\frac{\partial}{\partial t}w_{i}(t) & =-\frac{\partial}{\partial w_{i}}\,H(w,\mathcal{D})+\zeta_{i}(t)\\
 & =\sum_{\mu=1}^{P}\,\big(\phi(\bar{h}_{\mu})-\phi(h_{\mu})\big)\,\phi^{\prime}(h_{\mu})\,x_{\mu\,i}-\frac{1}{g\beta}w_{i}+\zeta_{i}(t)\,,\\
\langle\zeta_{i}(t)\zeta_{j}(s)\rangle & =\frac{2}{\beta}\,\delta_{ij}\,\delta(t-s)\,,
\end{align*}
so that the stationary distribution is given by $\exp\big(-\beta H(w,\cD)\big)$.
This equation of motion can be seen as an equation of motion on the
level of the pre-activations $h_{\nu}$ by defining $h_{\mu}:=w^{\T}x_{\mu}=\sum_{i=1}^{d}\,w_{i}x_{\mu i}$
as well as $\bar{h}_{\mu}:=\bar{w}^{\T}x_{\mu}$ and $\zeta_{\nu}(t):=\sum_{i=1}^{d}x_{\nu i}\,\zeta_{i}(t)$
to obtain
\begin{align*}
\frac{\partial}{\partial t}h_{\nu} & =\sum_{\mu=1}^{P}\,\big(\phi(\bar{h}_{\mu})-\phi(h_{\mu})\big)\,\phi^{\prime}(h_{\mu})\,\sum_{i=1}^{d}\,x_{\mu\,i}x_{\nu\,i}-\frac{1}{g\beta}h_{\nu}+\zeta_{\nu}(t)\,.
\end{align*}
Introducing the shorthands
\begin{align*}
C_{\mu\nu}^{(xx)} & :=\sum_{i=1}^{d}x_{\mu\,i}\,x_{\nu\,i}\,,\\
\delta\phi_{\mu}(h_{\mu}) & :=\phi(h_{\mu})-\phi(\bar{h}_{\mu})\,,\\
\Phi(h_{\mu}\,;\bar{h}_{\mu}) & :=\delta\phi_{\mu}(h_{\mu};\,\bar{h}_{\mu})\phi^{\prime}(h_{\mu})\,,
\end{align*}
the equation of motion takes the form
\begin{align}
\frac{\partial}{\partial t}h_{\nu} & =-\sum_{\mu=1}^{P}\,C_{\nu\mu}^{(xx)}\,\Phi(h_{\mu}\,,\,\bar{h}_{\mu})-\frac{1}{g\beta}h_{\nu}+\zeta_{\nu}(t)\,,\label{eq:eom_h}
\end{align}
which corresponds to the following moment generating functional
\begin{align*}
Z & =\int\cD h\,\int\cD\tilde{h}\,\exp\Big(\int dt\,\sum_{\nu=1}^{P}\,\tilde{h}_{\nu}(t)\,\left[\big(\partial_{t}+\frac{1}{g\beta}\big)\,h_{\nu}(t)+\sum_{\mu=1}^{P}\,C_{\nu\mu}^{(xx)}\,\Phi_{\mu}(t)\right]\\
 & +\frac{1}{\beta}\sum_{\mu,\nu=1}^{P}\tilde{h}_{\mu}(t)\,C_{\mu\nu}^{(xx)}\,\tilde{h}_{\nu}(t)\,\Big)\,.
\end{align*}
The disorder average over $C_{\mu\nu}^{(xx)}$, which again appears
linear in the exponent, can be treated in the same way as in the main
text and is left for future work. Note, though, that within the matrix
$C^{(xx)}$ the synapse index has been contracted, while in the linear
case considered in the main text, the contraction is over the sample
index. Still, both matrices are Wishart distributed and hence allow
for the same treatment.
\end{document}